\documentclass[preprint2]{aastex63}

\shorttitle{A Photometric Study of M67}
\shortauthors{Twarog, Anthony-Twarog, Deliyannis}

\begin{document}


\title{A $uvbyCa$H$\beta$ CCD Analysis of the Open Cluster Standard, M67, \\
and Its Relation to NGC 752}


\author{Bruce A. Twarog}
\affil{Department of Physics and Astronomy, University of Kansas, Lawrence, KS 66045-7582, USA}
\email{btwarog@ku.edu}

\author{Barbara J. Anthony-Twarog}
\affil{Department of Physics and Astronomy, University of Kansas, Lawrence, KS 66045-7582, USA}
\email{bjat@ku.edu}


\author{Constantine P. Deliyannis}
\affil{Department of Astronomy, Indiana University, Bloomington, IN 47405-7105 }
\email{cdeliyan@indiana.edu}





\begin{abstract}
Precision CCD $uvbyCa$H$\beta$ photometry is presented of the old cluster, M67, covering one square degree with typical internal precision at the 0.005-0.020 mag level to $V$ $\sim$ 17. The photometry is calibrated using standards over a wide range in luminosity and temperature from NGC 752 and zeroed to the standard system via published photoelectric observations. Relative to NGC 752, differential offsets in reddening and metallicity are derived using astrometric members, supplemented by radial-velocity information. From single-star members, offsets in the sense (M67 - NGC 752) are $\delta$$E(b-y)$ = -0.005 $\pm$ 0.001 (sem) mag from 327 F/G dwarfs and $\delta$[Fe/H] = 0.062 $\pm$ 0.006 (sem) dex from the combined $m_1$ and $hk$ indices of 249 F dwarfs, leading to $E(b-y)$ = 0.021 $\pm$ 0.004 (sem), and [Fe/H]$_{M67}$ = +0.030 $\pm$ 0.016 (sem) assuming [Fe/H]$_{Hyades}$ = +0.12. With probable binaries eliminated using $c_{1}, (b-y)$ indices, 83 members with $(\pi/\sigma_{\pi}) > 50$ generate $(m-M)_{0}$ = 8.220 $\pm$ 0.005 (sem) for NGC 752 and an isochronal age of 1.45 $\pm$ 0.05 Gyr. Using the same parallax restriction for 312 stars, M67 has $(m-M)$ = 9.77 $\pm$ 0.02 (sem), leading to an age tied solely to the luminosity of the subgiant branch of 3.70 $\pm$ 0.03 Gyr. The turnoff color spread implies $\pm$ 0.1 Gyr, but the turnoff morphology defines a younger age/higher mass for the stars, consistent with recent binary analysis and broad-band photometry indicating possible missing physics in the isochrones. Anomalous stars positioned blueward of the turnoff are discussed.

\end{abstract}


\section{Introduction}
Star clusters have long been extolled as critical testbeds of stellar evolution while simultaneously serving as well-defined, individual data points for probing the temporal, chemical, and spatial evolution of the Galaxy due to the unique distance, age, and chemical composition common to all stars within a cluster (see, e.g. \citet{TW97, FR02, NE16, DO20}). While the claim of uniformity for the latter two parameters has been successfully challenged by globular clusters exhibiting multigenerational abundance trends (see, e.g. \citet{GR04, CA09, MI12, PI20}), open clusters still retain the mantle of single-generation parametric homogeneity, apart from abundance variations due to the internal evolution of specific elements like Li or CNO. In particular, in the case of M67, diffusion and rotational mixing have arisen as a potential sources of significant multi-elemental variations with evolutionary phase \citep{SO19, BO20}. 

Of the thousands of clusters now known and isolated as physical entities within the Galactic environment, thanks in large part to the expanding astrometric insight supplied by the ongoing {\it Gaia} mission \citep{GA16, GA18, GA21, GA22}, with the possible exception of the very nearby Hyades, few open clusters have received as much attention as M67 with approximately 2200 published references, over one-third of these within the last decade. Its high profile was driven initially by: (a) a modest apparent distance modulus ($\sim$9.5-9.7) \citep{JO55, EG59, SA62, EG64}, reducing the excessive areal coverage necessary to compile a substantial cluster sample as required for nearby objects like the Hyades, NGC 752, and, more recently, Rup 147 \citep{CU13}; (b) an ``old" age 
($\sim$5-6 Gyr) \citep{SE69, VB85}, making it comparable to the sun while placing it with NGC 188 among the very few ``old" disk clusters accessible for probing the early evolution of the disk; (c) a commonly derived and adopted metallicity less than the Hyades and potentially similar, if not identical, to the sun \citep{EG64}; and (d) uniform and low reddening across the face of the populous cluster (see \citet{TA78} and references therein). Over the decades, significant and coupled changes have altered a number of the key cluster parameters, subtly impacting the contextual role of the cluster within stellar and galactic evolution. Compared to the initial cluster studies of 50 years ago, the current consensus, based upon a variety of analyses, places the cluster farther away with an age younger than the sun (see, e.g. \citet{SA21} (SA)). There is still no evidence for significant variation in reddening across the face of the cluster, but the absolute value of the reddening from some techniques exhibits offsets that measurably affect the cluster age estimate to an annoying degree in a time of supposedly precision photometry, astrometry, and stellar isochrones (see, e.g. \citet{TA07} and references therein). 

Of primary importance is the metallicity. The initial estimates for M67 from broad-band photometry and modest spectroscopy tagged the cluster as less metal-rich than the Hyades and potentially solar in metallicity (see \citet{TW78} and references therein), but the scatter in values from all techniques ranged from less than one-half solar \citep{CO80} to approximately twice the value of the Hyades \citep{SP69, SP70, GB72}. Fortunately, improvements in the quality and quantity of the photometric and spectroscopic analyses have reduced, but not eliminated, the range, with a current spread in [Fe/H] from just below solar to almost Hyades metallicity \citep{RE15, RA22}.

The aforementioned phase-dependent diffusion effects aside, potential sources of the metallicity offsets at the level of 0.05 to 0.1 dex among the various techniques for obtaining a mean cluster metallicity are numerous. To name just a few, for traditional spectroscopy the zero-point of the scale can be set by reference observations to one star, typically the Sun for dwarfs or a bright giant like Arcturus for evolved stars. Such approaches work well for stars with parameters ($T_{\rm{eff}}$ and log $g$) similar to the reference star but can become less reliable with increasing parametric distance from the standard. More recent approaches often make use of multiple spectroscopic standards or an array of synthetic spectra covering a range in [Fe/H], log $g$, and $T_{\rm{eff}}$, but the final values are only as reliable as the consistency of the adopted standard values or the accuracy of the atmospheric models. Different schemes for deriving the $T_{\rm{eff}}$, dependent for some techniques upon the adopted reddening and/or the microturbulent velocity, can generate small alterations in the metallicity zero-point, star-to-star scatter aside. When compounded with differences in spectral resolution, S/N, bandpass selection, and line lists, it is perhaps surprising that study-to-study comparisons of the same cluster don't show more variation.

For photometry, the observational approach is simpler and the transformation from photometric indices to stellar parameters is straightforward using relations defined by a large body of precision photometry of stars with independently derived fundamental parameters. Alternative relations linking observation to physical parameters may exist, but differences between these can be readily sorted to place any set of photometric stellar parameters on a common scale. Thus, the challenge for photometric abundance derivation in clusters is that of getting enough precision for individual stars in a large enough sample to reduce the cluster standard-error-of-the-mean (sem) to the desired precision. The potential sources affecting the zero-point accuracy of the metallicity determination bear some similarity to those of spectroscopy. Photometric indices are often designed to work well over a modest range in $T_{\rm{eff}}$, log $g$, and [Fe/H]; what supplies reliable abundances for giants may fail completely for dwarfs and vice versa. Derivation of photometric stellar parameters often requires correction for reddening, which may not be obtainable from the photometry itself. Invariably, the greatest uncertainty arises from the transformation of the photometry to the standard system; differences in photometric zero-points at the level of $\pm$0.01 mag for key indices, depending upon the photometric system, can generate metallicity offsets at the level of $\pm$0.1 dex or less. 

In the simplest terms, the purpose of the current investigation is to present precision photometry on the $uvbyCa$H$\beta$ system of a one-degree-square 
field which includes M67. The discussion follows the approach laid out in a similar survey of the nearby younger cluster, NGC 752 \citep{TW15} (hereinafter Paper I), with one key difference. While high quality photoelectric photometry will be used to set the zero-points of the M67 indices, the slopes of the calibration curves for all indices, giants and dwarfs independently when required, will be defined using the extensive CCD photometry of NGC 752 as presented in Paper I. This key cluster was observed during every run with M67 and supplies calibration standards for dwarfs and giants alike numbering in the hundreds, bypassing the lack of extensive standard star fields for intermediate-band photometry but common for traditional broad-band filters (see e.g. \citet{LA92, ST00}). The ultimate goals of this approach are three-fold: (1) with the photometry of both clusters on an internally coupled system across all temperatures and luminosities, highly reliable differential measures of the key cluster parameters of reddening, metallicity, distance, and age are attainable; (2) M67 has been used as a photometric link for precision observations of 5 additional clusters included in the ongoing survey of cluster Li abundances, all but one of which have no previous internal intermediate-band photometry. These data will allow the same differential approach used with NGC 752 to be applied to the less well-studied clusters covering a significant range in age and metallicity; (3) the fact that so much detailed analysis is available for the rich population of stars in M67 provides a testing ground for ways in which the multiple combinations of indices for either dwarfs or giants can be used to identify and isolate subclasses of stars of evolutionary interest for application to stellar systems where the dataset may be restricted to photometry alone. 

The outline of the paper is as follows: Section 2 discusses the collection and processing of the CCD observations of M67 while Section 3 details the procedure for compiling the internal photoelectric standards used to define the CCD zero-points, as well as the transformation of the instrumental data to the standard system using NGC 752 to generate the transformation slopes at all temperatures and luminosities. Section 4 uses {\it Gaia} astrometry and ground-based radial velocities (if available) to isolate probable single-star members and rederive the cluster reddening and metallicity using the precision photometry of Paper I. Section 5 applies the same membership approach to M67 and uses this select sample to derive the reddening and metallicity relative to NGC 752. Section 6 details the derivation of both distance and age for both clusters, identifying and exploring the discrepancies between theory and observation while Section 7 summarizes our conclusions.

\section{Observations and Data Reduction}
Intermediate and narrow-band images of M67 were obtained using the WIYN 0.9-m telescope during four observing runs between Nov. 2015 and Feb. 2017. During each run frames also were obtained in multiple fields of NGC 752, observed as the primary source of standards for the extended Str\"omgren and H$\beta$ systems. Because of the reduction approach outlined below, frames of both clusters were collected on both photometric and nonphotometric nights. For all runs the telescope was equipped with the Half-Degree-Imager (HDI), a 4K $\times$ 4K chip with 0.43\arcsec\ pixels covering a 29\arcmin $\times$ 29\arcmin\ field. The seven filters were from the extended Str\"omgren set acquired for specific use with the HDI.

Bias frames and dome flats were collected for every filter every night, while sky flats for all filters except $b$ and $y$ were obtained at twilight every night sky conditions allowed. Sky flats were always used in the frame processing except when the integrated counts on  the sky set of a given filter proved inadequate and the dome flats were adopted instead. To optimize the cluster frame collection, extinction fields were selected from the cluster fields themselves and monitored 3 or more times each night over a range in airmass, though for the current discussion, extinction corrections remain irrelevant.

To expand areal coverage, in contrast with the approach for NGC 752 (Paper I), M67 was divided into four barely overlapping fields, ultimately linked
through a central field overlapping with a quarter of each of the outer four fields. Given the 29\arcmin $\times$ 29\arcmin\ field of the HDI chip and the variable positioning of each field, the final photometry covers approximately one-square-degree. Exposure times in all filters were staggered to allow reliable photometry from $V$ $\sim$ 8.5 in all filters to varying depths in each. The total M67 frameset amounted to $\sim$450 frames over 7 filters in 5 fields.

A description of our procedures for processing and merging the photometry from multiple frames and fields is given in Paper I and will not be repeated. Suffice it to say that, unlike the earlier work with NGC 752, all frames for this investigation were collected with the same CCD chip and filter set, making the photometric merger both simpler and more reliable. Once transferred to a common coordinate system, all frames in a given filter were adjusted in magnitude using a derived offset to one frame of that filter adopted as the $standard$ for the instrumental system. Final instrumental magnitudes for each star were derived from the weighted averages of all frames for a given filter, with indices constructed from these averages. The final sem for each index is based upon the photometric scatter calculated from each magnitude used to construct the index. The sem for each star in $V$ and all five indices as a function of $V$ are plotted in  Figure 1. The sem is only calculated for stars with 3 or more observations in each filter used to construct an index. The tick marks on the vertical scale for all panels in Figure 1 define a change of 0.02 mag; the total range for sem is 0.10 mag for $V$ and 0.15 mag for the five photometric indices. As expected, due to the inclusion of a filter weighted toward the ultraviolet, $hk$ and $c_1$ indices have sem precision limits approximately 0.5 and 1.0 magnitude brighter than $m_1$, respectively.

\begin{figure}
\figurenum{1}
\includegraphics[angle=0,width=\linewidth]{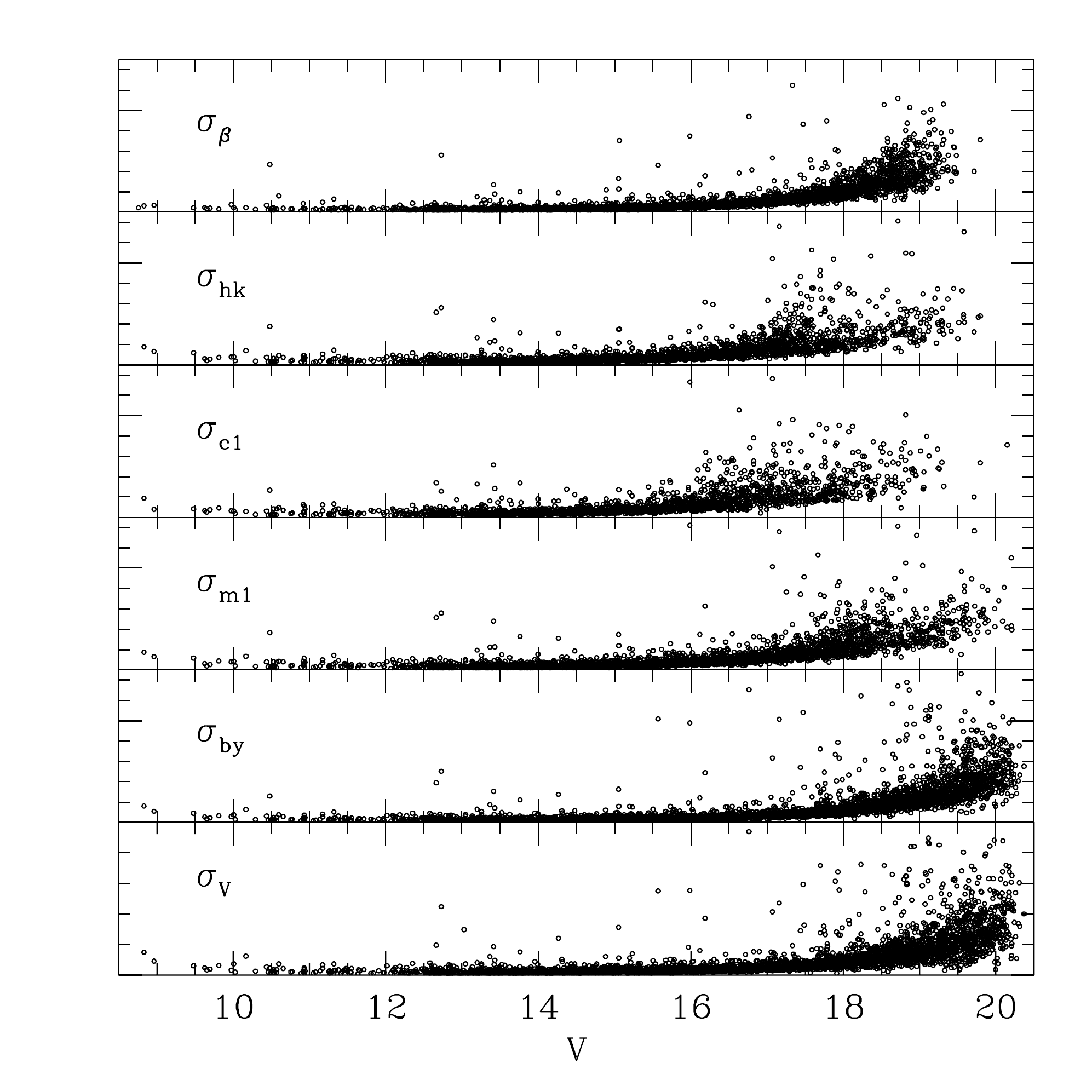}
\caption{Standard-errors-of-the-mean (sem) for each index and $V$ as a function of $V$ mag. The tick marks on the vertical scale for all panels define a 
change in the sem equal to 0.02 mag.}
\end{figure}

\section{Transforming the CCD Photometry}
A critical focus of current and future cluster comparisons is the need to ensure that all cluster photometry is on a common system, supplying 
confidence that differences among the indices between clusters are signatures of true differences in the relative cluster parameters rather than 
byproducts of calibration offsets. Fortunately, the two clusters of interest have been exhaustively investigated in a series of papers designed specifically to detect and minimize any zero-point differences in broad-band VRI \citep{JO08, TA08, TA11} and $uvby$H$\beta$ \citep{JO95, JO97} photometry. As with NGC 752 (Paper I), we will first compile a set of internal M67 $uvbyCa$H$\beta$ photoelectric standards tied to the established systems. Unlike NGC 752, however, these data will only be used to fix the zero-points for the transformations between the CCD instrumental system and the standard system once the transformation slopes have been defined using the NGC 752 CCD observations. This approach is crucial given that the internal $uvby$H$\beta$ standards for M67 include no red giants or red dwarfs and simple linear extrapolation of the relations from bluer stars, particularly for CCD photometry, does not work (Paper I). 

\subsection{Internal M67 Standards: $V$, $uvby$H$\beta$}
The $y$ magnitudes of the Str\"omgren system are unique in that, despite the narrower bandwidth of the filter compared to traditional $V$ filters, they can 
be transferred directly to the Cousins $V$ system with usually only a modest linear color-dependent correction. The obvious advantage is that, unlike the multifilter Str\"omgren indices, reliable broad-band photometry can be adopted to calibrate the $y$ magnitudes without the need to call solely upon $V$ defined through intermediate-band observations, photoelectric or otherwise. There have been many broad-band surveys of M67 over the years, beginning with the photoelectric data of \citet{EG64} through the photographic work of \citet{RA71}, to the CCD studies of \citet{MO93} and \citet{SA04}, among others. \citet{TA08} supply a comprehensive discussion of multiple sources of $V$ photometry for M67 on the Cousins system, producing two catalogs for the cluster, one based upon corrected data of \citet{SA04} alone (210 stars) and a second composed of a combination of recalibrated sets of both photoelectric and CCD data (241 stars). Because it covers a wider range in color though with a brighter magnitude limit, we will adopt the latter catalog to transfer our $y$ mags to the $V$ system, but use the former as a secondary check.

The largest set of $uvby$H$\beta$ photoelectric observations of M67 is that of \citet{NI87}(hereinafter NTC) which includes a mixture of indices for 79 stars, 33 of which have H$\beta$. Given the size and precision of the samples, our first step will be a merger with the data of \citet{JO95, JO97}, providing a reliable data set for testing and redefining the photometry from a number of smaller, less precise compilations, usually heavily weighted to the brighter blue stragglers that populate the cluster field. Comparisons will only be presented for data sets where the overlap between the sample and this photometric set of indices, referred to as the core set, is statistically significant enough to define a reliable offset calculation. In the following discussion, quoted uncertainties refer to the standard deviations among the residuals, unless otherwise noted.

The combined M67 H$\beta$ set of 22 stars from \citet{TA78} and \citet{JO97} has a 19 star overlap with NTC. Unweighted residuals, in the sense (JT-NTC), generate a mean offset of -0.005 $\pm$ 0.014. If one star with a significantly larger than average residual is dropped, the offset shifts slightly to -0.007 $\pm$ 0.011. Applying an offset of -0.006 mag to NTC, H$\beta$ photometry for stars in common to the two samples was averaged using a weighting by the inverse square of their individual errors, leading to a combined core H$\beta$ sample of 36 stars. 

The next H$\beta$ comparison is with the sample of 9 blue stragglers in \citet{EG81}, 8 of which overlap with our newly defined core. The derived offset, 
in the sense (CORE - EG), of -0.015 $\pm$ 0.008 mag has been applied to the \citet{EG81} data. 

The final attempted H$\beta$ match is to 21 stars of \citet{ST71}, 16 of which overlap with the core. The mean residual in H$\beta$, in the sense (CORE - SSB), is +0.006 $\pm$ 0.027 mag. Despite a larger overlap and expected improvement in the precision of the core sample, the rms scatter among the residuals is the same as that derived by NTC from their sample alone, implying that the primary source of the noise lies with the \citet{ST71} data. Because of the large photometric uncertainty implicit in their data, it was decided to exclude the \citet{ST71} data set from the final H$\beta$ merger.

The three reliably recalibrated data sets discussed above were averaged by weighting the individual H$\beta$ values by the inverse squares of their photometric uncertainties, producing a final sample of 37 stars. 

For $b-y$, $m_1$, and $c_1$, the core system is again defined by the merger of the NTC data with that of \citet{JO95, JO97}. Before discussing the offsets, it should be noted that for 5 blue stragglers, \citet{JO97} supply two sets of data for some indices, the first set obtained not later than 1987 and the second in 1996. The implication is that these stars exhibit statistically significant evidence for potential long-term variability and that the individual indices should be treated as intrinsically different measures from two widely separated time frames. For the 9 pairs of duplicate indices affecting 4 stars, we have adopted a simple average of the two values as the correct index for each star. For $b-y$, $m_1$, and $c_1$, the mean offsets from 15 stars in common, in the sense (JT - NTC), are -0.005 $\pm$ 0.009 mag, +0.006 $\pm$ 0.009 mag, and +0.002 $\pm$ 0.019 mag, respectively. The offsets for $b-y$ and $c_1$ are 
in excellent agreement with the analysis of \citet{JO97}. However, the offset derived for $m_1$ by \citet{JO97} is +0.0006 mag, significantly smaller than found here. As opposed to taking a simple average, we have attempted various combinations of the indices using different assumptions for which set of paired data to use for the blue stragglers, including total exclusion of the paired sets, and are unable to obtain an offset value as small as 0.001 mag; the full range of offsets goes from 0.004 mag to 0.009 mag. As with the H$\beta$ comparison, we will retain our calculated value as the appropriate offset. (For a detailed discussion of the many issues associated with the zero-point of the $m_1$ photoelectric system the reader is referred to the 
Appendix of NTC.)

With the NTC photometry adjusted and merged with that of \citet{JO95, JO97}, the first comparison is with the $uvby$ data from \citet{EG81}. Unlike the 
H$\beta$ data, the $m_1$ and $c_1$ indices are assumed to be on a slightly different system compared to the core data due to differences in the $v$ filter 
adopted by \citet{EG81}, thus leading to the identification of the indices as $M_1$ and $C_1$. For the 8 stars overlapping with the core, all blue stragglers, 
the mean residuals, in the sense (CORE - EG), are +0.004 $\pm$ 0.008 mag, +0.003 $\pm$ 0.004 mag, and -0.032 $\pm$ 0.023 mag 
for $(b-y)$, $m_1$, and $c_1$, respectively.

The next data set is that of \citet{BP71} for 7 blue stragglers, all of which overlap with the core sample. The mean offsets are -0.007 $\pm$ 0.013 mag, +0.012 
$\pm$ 0.020 mag, and +0.010 $\pm$ 0.016 mag for $(b-y)$, $m_1$, and $c_1$, respectively, in excellent agreement with the comparison in \citet{JO97}. 
Part of the scatter is due to the photometry being listed to only two decimal places.

Despite the apparent failure of the H$\beta$ comparison, a check was repeated for the $uvby$ set of \citet{ST71} using the 16 stars that overlap with 
the core. Previous attempts to transfer these data to a standard system have presented challenges, illustrated by the need to either break the sample into 
two distinct color ranges \citep{JO97} or include a color-dependent term within the offset (NTC). With the larger database provided by the core sample, 
it became apparent that both approaches were partially correct. For all three indices, the residuals show a well-defined pattern. For stars with $(b-y)$ 
below 0.31, there is a constant offset of -0.012 $\pm$ 0.009 mag, +0.015 $\pm$ 0.008 mag, and +0.038 $\pm$ 0.020 mag, respectively, for $(b-y)$, 
$m_1$, and $c_1$. For the stars redder than $(b-y)$ = 0.31, the offset includes a color term in $(b-y)$ with a slope of -0.29, 0.48, and -0.74 for $(b-y)$, $m_1$, and 
$c_1$, respectively. Application of these offset transformations produces photometry on the core system with a residual scatter of $\pm$0.009, $\pm$0.012, and $\pm$0.020 mag for $(b-y)$, $m_1$, and $c_1$, respectively.

As with H$\beta$, the core-recalibrated $(b-y)$, $m_1$, and $c_1$ indices from the five sources above were merged using the inverse square of the photometric uncertainties as 
weights, producing the M67 internal photoelectric standards compiled in Table 1. Identification numbers as defined in WEBDA are included, as well as the coordinates on the {\it Gaia} DR3 system. The other columns are self-explanatory. 

\floattable
\begin{deluxetable}{rrrrrrrrrrr}
\tablenum{1}
\tablecaption{Merged Photoelectric Secondary Standards in M67}
\tablewidth{0pt}
\tablehead{
\colhead{WEBDA ID} & \colhead{$\alpha (2000)$} & \colhead{$\delta (2000)$} & \colhead{$b-y$} & \colhead{sem} & \colhead{$m_1$} & 
\colhead{sem} & \colhead{$c_1$} & \colhead{sem} & \colhead{H$\beta$} & \colhead{sem} } 
\startdata
24 & 132.72147 & 11.79282 & 0.371 & 0.008 & 0.180 & 0.010 & 0.399 & 0.011 &  & \\
30 & 132.73199 & 11.87073 & 0.395 & 0.006 & 0.172 & 0.004 & 0.445 & 0.005 & 2.626 & 0.010 \\
43 & 132.74675 & 11.77027 & 0.369 & 0.012 & 0.226 & 0.014 & 0.365 & 0.015 &  & \\
48 & 132.75436 & 11.83635 & 0.462 & 0.007 & 0.215 & 0.009 & 0.388 & 0.010 &  & \\
54 & 132.76352 & 11.76316 & 0.391 & 0.002 & 0.184 & 0.002 & 0.433 & 0.003 & 2.649 & 0.020 \\
55 & 132.76459 & 11.75077 & 0.173 & 0.002 & 0.219 & 0.002 & 0.859 & 0.004 & 2.794 & 0.003 \\
61 & 132.77007 & 11.76579 & 0.399 & 0.007 & 0.177 & 0.009 & 0.363 & 0.009 &  & \\
64 & 132.77394 & 11.72970 & 0.434 & 0.009 & 0.199 & 0.011 & 0.349 & 0.011 &  & \\
73 & 132.78564 & 11.84806 & 0.358 & 0.006 & 0.188 & 0.008 & 0.364 & 0.008 &  & \\
75 & 132.78593 & 11.76993 & 0.369 & 0.007 & 0.179 & 0.008 & 0.397 & 0.009 &  & \\
77 & 132.78847 & 11.80572 & 0.367 & 0.005 & 0.171 & 0.005 & 0.386 & 0.006 &  & \\
79 & 132.78957 & 11.69583 & 0.466 & 0.007 & 0.225 & 0.009 & 0.417 & 0.010 &  & \\
80 & 132.79846 & 11.81404 & 0.364 & 0.006 & 0.189 & 0.008 & 0.380 & 0.008 &  & \\
81 & 132.79903 & 11.75613 & -0.035& 0.002 & 0.123 & 0.002 & 0.628 & 0.005 & 2.767 & 0.001 \\
82 & 132.80089 & 11.80964 & 0.375 & 0.007 & 0.163 & 0.009 & 0.376 & 0.009 &  & \\
83 & 132.80118 & 11.77257 & 0.376 & 0.004 & 0.180 & 0.004 & 0.404 & 0.004 & 2.609 & 0.005 \\
89 & 132.81010 & 11.84452 & 0.364 & 0.006 & 0.181 & 0.007 & 0.386 & 0.008 &  & \\
91 & 132.81145 & 11.78999 & 0.341 & 0.007 &       &       &       &       &  & \\
93 & 132.81265 & 11.82253 & 0.386 & 0.009 & 0.186 & 0.011 & 0.351 & 0.011 &  & \\
94 & 132.81387 & 11.83731 & 0.356 & 0.002 & 0.182 & 0.005 & 0.432 & 0.003 & 2.616 & 0.003 \\
98 & 132.81561 & 11.88300 & 0.356 & 0.001 & 0.184 & 0.010 & 0.415 & 0.012 &  & \\
103 & 132.82071 & 11.83594 & 0.331 & 0.008 &      &       &       &       &  & \\
106 & 132.82224 & 11.78351 & 0.341 & 0.008 & 0.194 & 0.010 & 0.394 & 0.011 &  & \\
111 & 132.82492 & 11.76505 &       &       &       &       &       &       & 2.630 & 0.007 \\
112 & 132.82536 & 11.71520 & 0.372 & 0.006 & 0.187 & 0.007 & 0.401 & 0.008 &  &  \\
\enddata
\tablecomments{(This table is available in its entirety in machine-readable and Virtual Observatory (VO) forms.)}
\end{deluxetable}

For $hk$, photoelectric photometry in M67 on the $ybCa$ system was obtained as part of the compilation of the original catalog of stars defining the fundamental system 
\citep{TA95}, though not included in the published catalog. Presented in Table 2 are the $(b-y)$, $hk$ data for 19 stars in the field of M67, ranging from blue stragglers through red giants. Identification of each star is via WEBDA number and (RA, Dec) coordinates on the {\it Gaia} DR3 system.

\floattable
\begin{deluxetable}{rrrrrrrr}
\tablenum{2}
\tablecaption{Internal Photoelectric $hk$ Secondary Standards in M67}
\tablewidth{0pt}
\tablehead{
\colhead{WEBDA ID} & \colhead{$\alpha (2000)$} & \colhead{$\delta (2000)$} & \colhead{$b-y$} & \colhead{sd} & \colhead{$hk$} & 
\colhead{sd} & \colhead{$n$}  } 
\startdata
81 & 132.79903 & 11.75613 & -0.029 & 0.008 & 0.144 & 0.008 & 2 \\
84 & 132.80284 & 11.87844 & 0.676 & 0.009 & 1.180 & 0.015 & 4 \\
105 & 132.82120 & 11.80448 & 0.771 & 0.006 & 1.460 & 0.030 & 2 \\
108 & 132.82281 & 11.75630 & 0.842 & 0.007 & 1.630 & 0.019 & 8 \\
127 & 132.83380 & 11.77825 & 0.367 & 0.005 & 0.571 & 0.011 & 3 \\
134 & 132.83838 & 11.76464 & 0.367 & 0.006 & 0.599 & 0.013 & 3 \\
135 & 132.83980 & 11.76838 & 0.649 & 0.004 & 1.167 & 0.009 & 3 \\
141 & 132.84497 & 11.80048 & 0.671 & 0.006 & 1.176 & 0.021 & 2 \\
151 & 132.85905 & 11.89776 & 0.664 & 0.000 & 1.208 & 0.000 & 1 \\
164 & 132.87075 & 11.84252 & 0.681 & 0.000 & 1.205 & 0.000 & 1 \\
170 & 132.87456 & 11.78800 & 0.830 & 0.003 & 1.596 & 0.025 & 3 \\
223 & 132.93281 & 11.94513 & 0.670 & 0.016 & 1.186 & 0.012 & 2 \\
231 & 133.04323 & 12.04647 & 0.651 & 0.000 & 1.129 & 0.000 & 1 \\
244 & 132.95913 & 11.76858 & 0.580 & 0.007 & 0.943 & 0.012 & 2 \\
266 & 132.99796 & 11.91800 & 0.668 & 0.000 & 1.167 & 0.000 & 1 \\
286 & 133.07734 & 11.74065 & 0.659 & 0.006 & 1.238 & 0.026 & 2 \\
2152 & 133.04566 & 11.53033 & 0.692 & 0.000 & 1.347 & 0.000 & 1  \\
6469 & 132.39436 & 11.85713 & 0.835 & 0.001 & 1.611 & 0.017 & 2 \\
6515 & 133.06894 & 11.32727 & 0.774 & 0.003 & 1.499 & 0.023 & 2 \\
\enddata
\end{deluxetable}

\subsection{Defining the Calibration Relations}
To define the slopes and/or color terms for the transformation of 
the instrumental extended Str\"omgren data to the standard system, use was made of the NGC 752 frames 
obtained and compiled during the same observing runs as M67. Because of the extensive set of frames covering all the fields of Paper I, almost 1770 stars 
brighter than $V$ = 18 were cross-matched with the final indices of Paper I. To optimize the precision of the calibration, calibration stars were retained 
only if there were 3 or more observations in each filter used to construct an index for both the standard and the instrumental data. To improve the 
calibration definition, cuts were also made based upon the calculated sem for both the instrumental and standard photometry. 
These limits will be detailed within the discussion of the individual indices.

For all indices and the $V$ magnitude for the standard stars in NGC 752, a general calibration equation of the form 

\noindent
INDEX$_{stand}$ = $a$*INDEX$_{instr}$ + $b*(b-y)_{instr}$ +  $c$

\noindent
was adopted. Following the procedure outlined in Paper I, for $V$ and $hk$ stars of all luminosity classes were treated as a single group. For the other indices the sample was separated into three groups: cooler dwarfs, blue dwarfs, and red giants, as defined in Table 4 of Paper I. Calibration relations were tested both individually and in combination for the three categories and the optimal fit adopted for each index. The resulting calibration slopes, {\it a} and {\it b}, along with the number of stars used in each calibration, are 
listed in Table 3. We will discuss the definition of the zero-points, {\it c}, in the next subsection.

\floattable
\begin{deluxetable}{rrrrrrrrr}
\tablenum{3}
\tablecaption{Summary of Transformation Coefficients}
\tablewidth{0pt}
\tablehead{
\colhead{Index} & \colhead{Class} & \colhead{$a$} & \colhead{$b$} & \colhead{$c$} & 
\colhead{$N_{752}$} & \colhead{$RES_{752}$} & \colhead{$N_{M67}$} & \colhead{$RES_{M67}$} } 
\startdata
$V$ & All & 1.000 & 0.070 & 1.391 &  &  & 217 & 0.010 \\
$hk$ & All & 1.161 & 0.000 & -2.027 & 597 & 0.019 & 19 & 0.020 \\
H$\beta$ & RG/BD & 1.092 & 0.000 & 0.471 & 451 & 0.011 & 35 & 0.010 \\
H$\beta$ & RD & 1.000 & 0.000 & 0.637 & 121 & 0.014 &  &  \\
$b-y$ & RG/BD & 1.060 & 0.000 & 0.208 & 648 & 0.011 & 74 & 0.010 \\
$b-y$ & RD & 0.900 & 0.000 & 0.248 & 221 & 0.012 &  &  \\
$m_1$ & BD & 1.000 & 0.000 & -1.039 & 418 & 0.016 & 68 & 0.011 \\
$m_1$ & RG & 0.695 & 0.000 & -0.651 & 157 & 0.018 &  &  \\
$m_1$ & RD & 1.000 & 0.441 & -1.160 & 143 & 0.017 &  &  \\
$c_1$ & RG & 1.000 & 0.312 & 0.340 & 98 & 0.024 &  &  \\
$c_1$ & BD/RD & 1.058 & 0.000 & 0.403 & 442 & 0.027 & 65 & 0.016 \\
\enddata
\tablecomments{BD designates Blue Dwarfs; RD Red Dwarf; RG Red Giant.}
\tablecomments{Form of calibration: $INDEX_{stand}  = a \times INDEX_{instr} + b \times (b-y)_{instr} +  c $}
\end{deluxetable}

As noted earlier, for $V$ we make primary use of the \citet{TA08} compilation from a mixture of photoelectric and CCD data, bypassing the need to 
separately define the slope and zero-point of the photometric system. Having transferred our (X,Y) coordinates to the (RA, Dec) system defined by DR3, 
we cross-matched our photometry with the catalog of 241 stars from \citet{TA08}, identifying 239 stars in common. Of these, 6 did not have $V$ mags, 
only $RI$, and were dropped. Using the remaining 233 stars and assuming {\it a} = 1.00 in the calibration relation, an initial linear fit was made to the data to 
define the color slope and zero-point. All stars with residuals greater than 0.1 mag relative to the mean relation were removed and the process repeated. 
With a revised estimated scatter about the mean relation ($\sigma$) calculated, all stars with residuals greater than 3.5$\sigma$ were eliminated 
and the linear fit rederived. The process rapidly converged to a stable $\sigma$ = $\pm$0.010 and all stars with residuals greater than 0.035 mag eliminated, leaving a 
net of 217 standards and {\it b$_V$} $= 0.070 \pm 0.006$ (sem) and {\it c$_V$}$ = 1.391 \pm 0.002$ (sem). Of the 16 stars eliminated due to their residuals, 
10 were more than 0.05 mag removed from the mean relation. 
 
While the standard set used above was adopted because of its wide range in color, extending from extreme blue stragglers to cool red giants, it is also dominated 
by stars brighter than $V$ = 15. As a simple check, we also derived a calibration curve using 207 stars from the modified photometry of \citet{SA04}, eliminating 
7 stars due to larger than average residuals. As expected, the transformation relations from both catalogs are 
statistically indistinguishable, with the modified \citet{SA04} comparison showing residuals with larger scatter ($\pm$0.020) due to a range in $V$ 
extending to 18.7 among the standards.

For the $(b-y)$ calibration, in addition to eliminating stars with fewer than 3 observations in either filter for both the instrumental and standard system, 
stars with the sem for $(b-y)$ \textgreater\ 0.015 mag for either data set were eliminated. This left 863 potential standards at all colors and luminosity classes. 
Optimal fits between the instrumental and standard system were produced by dividing the sample into two categories: blue dwarfs through giants (650 stars) 
and red dwarfs (222 stars). Removal of 2 (1) anomalously deviant points for the blue dwarf/giant (red dwarf) data led to dispersions of the residuals around 
the mean relations (Table 3) amounting to $\pm$ 0.011 (0.012) mag.

Unlike the other indices, no distinction is made among the giants and dwarfs, red or blue, for defining the $hk$ calibration curve. Limiting our sample to all stars with at least 3 observations each for $b$, $y$, and $Ca$ for both standards and instrumental values and an sem limit for $hk$ of 0.020 generates a sample of 607 stars. 
After a preliminary fit to the data, 10 stars with anomalously large residuals were removed, leading to the calibration slopes of Table 3 and a dispersion of the residuals 
about the mean relation of $\pm$0.019 mag.

As discussed in Paper I, breaking the calibration sample for H$\beta$ into three distinct temperature/luminosity categories can be challenging given the modest range for the index across all temperatures ($\sim$2.45 to 2.95) coupled with the location of the boundary between cool dwarfs/giants and blue dwarfs near $(b-y)$ = 
0.45. This has an approximate location in H$\beta$ of 2.59. As illustrated in Paper I, the red giant H$\beta$ range extends to only 2.55, while the cool dwarf 
boundary defines the lower limit of the index near 2.45. Thus, any defined linear transformation of the giants alone is readily dominated by the photometric scatter 
within the index, equivalent in size to the full range of the index itself. For the current calibration, the red giants and blue dwarfs were transformed as one group while the red dwarfs were treated separately.

Eliminating all stars with instrumental and/or standard errors \textgreater\ 0.015 mag generated a sample of 464 blue dwarf/red giant stars. After a preliminary fit, removal of 13 stars with residuals larger than 0.03 mag led to a final calibration relation defined by 451 stars with a dispersion about the mean relation of $\pm$0.011 mag. For the red dwarfs, the difference in slope compared to the blue dwarf/red giant relation was immediately apparent. Applying the same internal error cut to the dwarfs resulted in a sample of 123 stars. A linear fit produced a slope statistically indistinguishable from 1.00.  Removing 2 stars with large residuals and adopting a 
slope, {\it a}, of 1.0, the scatter among the residuals is $\pm$0.014 mag. 

Unlike H$\beta$, among the three stellar classes $m_1$ has the smallest range among the blue dwarfs, despite a significant range in $(b-y)$. 
Restricting the sample to 419 stars with both instrumental and standard errors below 0.020 mag, one finds no statistically significant evidence for a color 
term, i.e. {\it b} = 0.000, or a slope, {\it a}, other than 1.0 linking the instrumental and standard systems. Eliminating three stars with anomalously large 
residuals produces a scatter among the residuals of $\pm$0.016 mag. For red dwarfs, the slope {\it a} is also 1.00, but a significant color term, {\it b}, now
emerges. From 143 red dwarfs, eliminating none due to large residuals, produces a scatter about the mean relation of $\pm$0.017 mag.
For the red giants, with a significantly greater range in $m_1$ and $(b-y)$, elimination of 9 giants with anomalous residuals leaves 157 stars. One derives 
the calibration relations of Table 3 with a residual scatter about the mean relation of $\pm$0.018 mag.        

Finally, for $c_1$, all stars with internal errors above $\pm$0.020 mag in either the standard or instrumental systems were eliminated. Calibrating all dwarfs, red and blue, with a common relation from 442 stars (Table 3), 3 eliminated due to excessive residuals, resulted in a scatter about the mean relation of $\pm$0.027 mag. For the red giants alone, elimnating 7 stars due to excessive residuals, produced a scatter of $\pm$0.024 mag from 98 stars.

\subsection{Zeroing the Scales}
While the slopes of the calibration relations are best set by the large database of NGC 752 photometry covering all luminosity classes, ideally the zero-points of the M67 CCD photometry 
should be linked to the $uvbyCa$H$\beta$ standard system via the well-defined photoelectric secondary standards presented in Tables 1 and 2. After applying the calibration relations to the CCD data for the standards of Tables 1 and 2, the zero-points for each index were then derived by minimizing the residuals between the two systems. We note again that, unlike $V$ and $hk$, with the exception of star 117 which lies just beyond the edge of the blue dwarf boundary, all M67 photoelectric standards in $(b-y)$, $m_1$, $c_1$, and H$\beta$ fall within the category of blue dwarfs. Thus the zero-points for the red dwarfs and the red giants for these indices are derived indirectly through the relative relationships these three classes define among the NGC 752 sample. Table 3 lists the numbers of stars used in defining the zero-points, 
as well as the scatter among the residuals for each index. Three stars (115, 131, 132) exhibited anomalously large residuals in $c_1$ and were excluded from the final determination of the zero-point for that index. Since it is probable that in these cases the issue lies with the photoelectric data, these stars have been flagged with note in Table 1 indicating that the values should be treated with caution. For the case of star 155, a blue straggler, the instrumental indices showed signs of variability. This star was excluded from the zero-point determinations, indicated by a flag for this star in Table 1.

The final M67 photometry is presented in Table 4, where the columns are self-explanatory. Coordinates are on the \citep{GA22} (DR3) system. With the exception of a few stars above $V$ = 14 where the internal errors for the frames are assumed to be small due to the brightness of the stars, photometric indices are only listed if every filter within an index has at least 3 observations. If not, the number of filter observations has been set to 0, with the index and error values set to a null indicator value of 9.999.
 As in Paper 1, the final column of the Table lists the categorization of the star as a blue dwarf, red dwarf, or red giant. Unlike Paper I, the ability to separate red dwarfs and red giants has been greatly enhanced by the availability of {\it Gaia} parallaxes. For the subset of stars for which parallax is either unavailable or poorly determined, use has been made of the photometric indices, following the pattern of Paper I. For fainter stars with questionable parallax and potentially unreliable photometric indices, it has been assumed that any red star is highly likely to be a red dwarf rather than an exceptionally distant red giant situated well above the galactic plane ($b$ = +32$\arcdeg$).

\floattable
\begin{deluxetable}{rrrrrrrrrrrrrrrrrrrrrr}
\tablenum{4}
\tablecaption{Str\"omgren Photometry in M67}
\tabletypesize\tiny
\tablewidth{0pc}
\setlength{\tabcolsep} {0.04in}
\tablehead{
\colhead{$\alpha(2000)$} & \colhead{$\delta(2000)$} & \colhead{$V$} & \colhead{$b-y$} & \colhead{$m_1$} & 
\colhead{$c_1$} & \colhead{$hk$} & \colhead{H$\beta$} & \colhead{$sem(V)$} & \colhead{$sem(by)$} & 
\colhead{$sem(m1)$} & \colhead{$sem(c1)$} & \colhead{$sem(hk)$} & \colhead{$sem\beta$}  & \colhead{$N_y$} & 
\colhead{$N_b$} & \colhead{$N_v$} & \colhead{$N_u$} & \colhead{$N_Ca$} & \colhead{$N_n$} & 
\colhead{$N_w$} & \colhead{Class} }
\startdata 
132.57755 & 11.40783 & 8.758 & 0.623 & 0.452 & 0.380 & 1.278 & 2.568 & 0.002 & 0.005 & 0.009 & 0.010 & 0.008 & 0.004 & 1 & 1 & 6 & 10 & 10 & 7 & 5 & G \\
132.55121 & 11.85679 & 8.830 & 0.998 & 0.668 & 0.271 & 1.954 & 2.576 & 0.015 & 0.016 & 0.017 & 0.019 & 0.018 & 0.006 & 3 & 3 & 14 & 5 & 15 & 8 & 8 & G \\
132.50999 & 11.92349 & 8.961 & 0.670 & 0.549 & 0.022 & 1.216 & 2.569 & 0.009 & 0.011 & 0.013 & 0.008 & 0.013 & 0.007 & 3 & 3 & 10 & 13 & 11 & 8 & 8 & G\\
132.39436 & 11.85713 & 9.481 & 0.853 & 0.593 & 0.328 & 1.607 & 2.556 & 0.006 & 0.009 & 0.012 & 0.008 & 0.012 & 0.004 & 3 & 5 & 7 & 5 & 8 & 4 & 5 & G \\
132.68250 & 12.12799 & 9.626 & 1.047 & 0.634 & 0.296 & 2.002 & 2.586 & 0.005 & 0.005 & 0.006 & 0.007 & 0.006 & 0.005 & 4 & 7 & 14 & 6 & 14 & 10 & 8 & G \\
132.87456 & 11.78800 & 9.654 & 0.822 & 0.589 & 0.330 & 1.607 & 2.562 & 0.003 & 0.004 & 0.004 & 0.005 & 0.004 & 0.004 & 20 & 19 & 29 & 20 & 35 & 27 & 18 & G \\
132.82281 & 11.75630 & 9.696 & 0.844 & 0.600 & 0.344 & 1.656 & 2.565 & 0.004 & 0.005 & 0.005 & 0.008 & 0.006 & 0.004 & 19 & 20 & 28 & 17 & 36 & 26 & 21 & G \\
132.48669 & 11.69248 & 9.813 & 0.807 & 0.590 & 0.319 & 1.592 & 2.577 & 0.006 & 0.007 & 0.009 & 0.009 & 0.007 & 0.003 & 5 & 5 & 10 & 7 & 13 & 8 & 7 & G \\
132.57610 & 11.92254 & 9.976 & 0.639 & 0.426 & 0.398 & 1.209 & 2.546 & 0.003 & 0.006 & 0.008 & 0.006 & 0.008 & 0.008 & 5 & 7 & 16 & 22 & 20 & 15 & 12 & G \\
132.45908 & 11.42578 & 10.008 & 0.568 & 0.335 & 0.396 & 1.012 & 2.581 & 0.007 & 0.008 & 0.009 & 0.006 & 0.008 & 0.005 & 6 & 7 & 11 & 12 & 13 & 8 & 10 & G \\
132.79903 & 11.75613 & 10.026 & -0.040 & 0.112 & 0.640 & 0.130 & 2.781 & 0.002 & 0.003 & 0.004 & 0.004 & 0.004 & 0.003 & 20 & 14 & 15 & 17 & 23 & 23 & 16 & B \\
133.06894 & 11.32727 & 10.053 & 0.807 & 0.549 & 0.343 & 1.493 & 2.551 & 0.002 & 0.003 & 0.004 & 0.010 & 0.006 & 0.006 & 2 & 2 & 3 & 2 & 3 & 3 & 2 & G \\
132.83279 & 11.36793 & 10.167 & 0.590 & 0.376 & 0.385 & 1.035 & 2.561 & 0.012 & 0.013 & 0.014 & 0.006 & 0.014 & 0.005 & 8 & 9 & 16 & 18 & 22 & 17 & 12 & G \\
132.82120 & 11.80448 & 10.293 & 0.768 & 0.544 & 0.358 & 1.451 & 2.567 & 0.003 & 0.003 & 0.003 & 0.003 & 0.004 & 0.003 & 33 & 31 & 42 & 40 & 51 & 39 & 40 & G \\
133.07735 & 11.74065 & 10.438 & 0.665 & 0.448 & 0.368 & 1.179 & 2.547 & 0.005 & 0.006 & 0.008 & 0.006 & 0.007 & 0.007 & 12 & 13 & 16 & 15 & 20 & 15 & 13 & G \\
132.84497 & 11.80048 & 10.458 & 0.668 & 0.453 & 0.376 & 1.203 & 2.564 & 0.002 & 0.002 & 0.003 & 0.002 & 0.003 & 0.002 & 46 & 41 & 56 & 58 & 66 & 54 & 48 & G \\
132.86166 & 11.81121 & 10.480 & 0.373 & 0.180 & 0.359 & 0.632 & 2.607 & 0.005 & 0.026 & 0.037 & 0.027 & 0.038 & 0.047 & 3 & 6 & 13 & 16 & 17 & 10 & 5 & B \\
132.85905 & 11.89776 & 10.493 & 0.657 & 0.440 & 0.385 & 1.182 & 2.562 & 0.002 & 0.002 & 0.003 & 0.002 & 0.003 & 0.002 & 35 & 30 & 35 & 35 & 42 & 34 & 31 & G \\
132.99796 & 11.91800 & 10.509 & 0.663 & 0.433 & 0.389 & 1.169 & 2.562 & 0.003 & 0.003 & 0.004 & 0.003 & 0.004 & 0.004 & 25 & 20 & 21 & 22 & 27 & 23 & 23 & G \\
132.80284 & 11.87844 & 10.524 & 0.665 & 0.440 & 0.397 & 1.196 & 2.572 & 0.002 & 0.002 & 0.003 & 0.003 & 0.003 & 0.002 & 20 & 21 & 24 & 24 & 29 & 23 & 23 & $G$ \\
\enddata
\tablecomments{(This table is available in its entirety in machine-readable and Virtual Observatory (VO) forms.)}
\end{deluxetable}
\twocolumngrid

CCD observations of M67 on the Str\"omgren and H$\beta$ systems are few and far between. The first attempt to test the use of a CCD to obtain $uvby$ photometry was made by 
\citet{AT87} on M67. The data included only two frames in each color for two 3\arcmin $\times$ 5\arcmin\ fields taken with the 4-m Blanco telescope, non-imaging $uvby$ 
filters and a high-readout-noise, low-$u$ sensitivity RCA chip. Not surprisingly, the $uvby$ data calibrated directly to the internal standards of NTC produced very similar results for the cluster parameters, though with double the uncertainty. 
 
More recently, \citet{BA07} have attempted an evaluation of the fundamental M67 cluster parameters from wide-field CCD $uvby$H$\beta$ photometry. While the quoted internal precision is comparable to the present sample for stars $V \leq 17$ for most indices,  H$\beta$ data are available for less than 20\% of the sample and are  
significantly less accurate. Equally problematic, the photometry was calibrated solely using NTC despite the lack of cool dwarf and red giant standards. Membership was estimated 
from ground-based astrometry and photometric criteria, isolating 776 members and, as with \citet{AT87}, produced distance moduli and metallicities the same, within the large errors, as NTC.

\section{NGC 752 Revisited}
\subsection{Core Cluster Membership}
The initial step in the modern evaluation of any open cluster sample is the detection and isolation of probable cluster members via the astrometric database supplied by {\it Gaia}, the current version being the DR3 data release. For NGC 752, the transition from the definitive ground-based proper-motion study by \citet{PL91}(PL) to the initial membership survey by \citet{CA18} using both proper motion ($\mu$) and parallax ($\pi$) expanded the cluster database in both magnitude range and sky coverage, while significantly improving the astrometric precision. Until very recently, the third kinematic component, radial velocity, was only available for a modest subset of the cluster sample, dominated early on by the precision ground-based observations of \citet{ME09}, but expanded slowly by more recent spectroscopic surveys of the cluster. The DR3 database includes a more comprehensive set covering the entire field of the cluster, but the precision is modest for stars on the unevolved main sequence, making membership and binarity estimation problematic.

Of immediate interest, revisiting some of the pre-{\it Gaia} discussion of Paper I from an astrometric standpoint serves two purposes. First, the fundamental cluster parameters of reddening and metallicity were derived from photometric analysis of 68 highly probable cluster members, F dwarfs with proper motions and/or radial velocities consistent with the cluster average and no indications of photometric anomalies, i.e. variability or a deviant photometric metallicity. Despite the already high precision of these parameters, for consistency, any newly identified nonmembers and/or binaries should be eliminated from the averages. Second, and more important, while one can argue about the zero-point accuracy of the absolute parametric scales rather than their precision, our primary interest lies with differential cluster-to-cluster comparisons possible with high precision intermediate-band photometry when the program cluster (M67) has been photometrically calibrated using the reference cluster as the standard, rather than tying the program cluster into the standard system using a small sample of field stars of inadequate temperature and luminosity range. With the photometric zero-points and calibration curves optimized, one can derive more precise and accurate differential (cluster to cluster) reddening and metallicity, thereby hopefully leading to improved relative ages and, independent of the parallax, relative distances.

As a starting point we compiled two datasets. The 1590 stars of Table 4 in Paper I were matched with the coordinates of DR3, resulting in cross-identification of 1585 stars to $V$ = 18.0, keeping in mind that only $V, (b-y)$ data are available for all stars to this limit. The 253 cluster members as derived by \citet{CA18} were identified in DR3 and restricted to the same core area as Table 4 of Paper I, generating a preliminary core membership set of 145 stars. The reduced fraction of members relative to \citet{CA18} is readily explained by the difference in areal coverage of the two surveys ($\sim$0.75$\arcdeg$ x 0.75$\arcdeg$ for Paper I vs. 6.1$\arcdeg$ x 4.5$\arcdeg$ for \citet{CA18}) and an approximately one magnitude difference in the photometric limits of the two surveys. A simple check also shows that the DR3 data set has produced a tighter astrometric profile for the cluster. The core cluster members from \citet{CA18} had mean values of $\pi$, $\mu$$_{\alpha}$ and $\mu$$_{\delta}$ of 2.223 $\pm$ 0.073 mas, 9.784 $\pm$ 0.303 mas-yr$^{-1}$, and -11.686 $\pm$ 0.305 mas-yr$^{-1}$, respectively. The same stars within DR3 have the analogous values of 2.263 $\pm$ 0.053 mas, 9.723 $\pm$ 0.253 mas-yr$^{-1}$, and -11.806 $\pm$ 0.253 mas-yr$^{-1}$.

To isolate cluster members within the core, all stars with Str\"omgren photometry and $\mu$$_{\alpha}$ and $\mu$$_{\delta}$ within 0.759 mas-yr$^{-1}$ (3$\sigma$) of the cluster mean $\mu$$_{\alpha}$ and $\mu$$_{\delta}$ were identified. Additionally, all stars whose individual quoted uncertainties for each $\mu$ measure were within 3$\sigma$ of these boundaries were retained. The same test was then applied to $\pi$ and $\sigma_{\pi}$ for this restricted sample, leading to a final sample of 126 probable members in the cluster core with some degree of Str\"omgren photometry.

More recently, \citet{BH21} have revisited the entire cluster membership issue using the early-release \citep{GA21}(EDR3) dataset within a 5$\arcdeg$ radius of the cluster center. Due to the extended area of this survey, required to identify the tidal tails of the kinematically evaporating cluster, ML-MOC \citep{AG21}, a $k$-nearest neighbor algorithm coupled with a Gaussian mixture model, was used to isolate cluster members in $\mu$-$\pi$ space, resulting in 282 likely members. The larger sample compared to \citet{CA18} is primarily due to the addition of stars below $G$ = 18. For the cluster core data of Paper I, the \citet{BH21} analysis identifies 107 members compared to 110 from \citet{CA18}. The small difference in the overlap between the two datasets emerged as a byproduct of the larger than average astrometric errors among a handful of stars in the original membership analysis. Adoption of the \citet{BH21} core members as the basis for identifying and restricting the final DR3 membership among stars with Str\"omgren photometry produces essentially identical overlap with the 126 stars identified above.

To constrain the cluster membership further, we make use of the third kinematic component, the radial velocity. In addition to eliminating field star contamination, adequate sampling can also identify binaries, members which should be eliminated from intercluster comparisons to minimize distortions caused by photometric anomalies and/or non-standard evolution. For NGC 752, the baseline survey among the brighter stars is that of \citet{ME09}, where 55 of the 126 astrometric members have radial-velocity information. (In the discussion that follows, if identification numbers are given, they are from the proper-motion survey of PL.) Of the 55, one (772) is a definite radial-velocity nonmember and 15 are definite binaries. The remaining 39 are either definitely single or have radial velocities consistent with cluster membership but an inadequate sample to test for binarity. 

The next relevant survey is that of \citet{AG18} who made use of unpublished long-term, radial-velocity monitoring of stars in NGC 752 to identify 11 nonmembers, 6 of which appear within our original set of 126. One of these is the previously identified nonmember (772), so five others can be eliminated (477, 641, 728, 888, 1008). One star (786) has a radial velocity consistent with membership but is a definite binary. Among the probable proper-motion members at the time, \citet{AG18} confirmed the binarity of three stars within our sample (814, 857, 1117), as well as a binary nature for star 849, now classed as a proper-motion nonmember from {\it Gaia} data \citep{CA18}, a result confirmed with the more recent data releases \citep{RA21}. This last star is unique in that it had long been considered the sole blue straggler member of the cluster and analyzed even recently as such \citep{LE21}.

Turning next to \citet{MA13}, based on a single epoch of observation, 8 potential non-members or radial velocity variables were identified out of a sample of 45 stars.  
Four of these are also {\it Gaia}-based nonmembers. One, star 552, is an astrometric member and spectroscopic binary \citep{ME09}, thereby explaining its discrepant velocity. The remaining 3 (828, 964, 1161) are astrometric members and 964 is a definite radial-velocity member according to \citet{ME09}. Star 828 sits 4.3 km-sec$^{-1}$ ($\sim 5 \sigma$) above while 1161 is 7.5 km-sec$^{-1}$ ($\sim 9 \sigma$) below the cluster mean as derived by \citet{MA13}. An average of the DR3 radial velocities for the brightest stars with the smallest radial velocity uncertainties leads to a cluster mean of 5.4 $\pm$ 0.15 (sem) km-sec$^{-1}$. The DR3 radial velocities for these two stars place them both below the cluster 
mean by 3 and 9 km-sec$^{-1}$, respectively. Unfortunately, the uncertainties in the velocities are 4.7 and 5.5 km-sec$^{-1}$, respectively.  All three stars are assumed to be either binaries like 552 or nonmembers.

Stars 897 and 950 are QX And \citep{QI07} and DS And \citep{MI19}, respectively, two eclipsing binary members of NGC 752. The latter star is one of two eclipsing binaries in NGC 752 analyzed in detail by \citet{SA23}, a followup to their discussion of a detached binary system in M67, a star which plays a key role in the discussion of Section 6.

Finally, while supplying no radial velocities for their sample, \citet{LU19} did evaluate rotation speeds as part of the line measurement process. Two stars that were excluded from their analysis due to significant line broadening were 552 and 413. As noted above, the former star has been previously classed as a spectroscopic binary and \citet{LU19} apply the same description to both stars. The DR3 radial velocity for this star is 8.16 $\pm$ 0.97 km-sec$^{-1}$.

Removing all nonmembers and/or probable binaries leaves a sample of 98 stars; we emphasize again the lack of precision radial-velocity measures for stars $V \geq 14.0$. For these stars the position in the color-magnitude diagram (CMD) can offer some indication of binarity, a point we will return to below.

\subsection{Zeroing the Scale: Metallicity and Reddening for NGC 752}
The derivation of reddening and metallicity for NGC 752 based upon the extended Str\"omgren photometric system from precision CCD photometry of the cluster core is discussed in exceptional detail in Paper I and will not be repeated here. Suffice it to say that the reddening is defined by comparison of cluster F dwarfs in the $(b-y)$ - H$\beta$ plane to a standard sequence defined by nearby field stars with apparently zero reddening, and metallicity comparable to that of the Hyades. Since H$\beta$, as a line filter ratio, is designed  to be unaffected by reddening and exhibits minimal, if any, dependence upon metallicity, displacements in $(b-y)$ from the standard relation are assumed to be signatures of differences in reddening and/or metallicity. With reddening determined, the metallicity-dependent indices of $m_{1}$ and $hk$ can be adjusted for reddening and metallicity derived independently from comparison of each index to a standard, dereddened relation tied to an adopted Hyades-metallicity sequence. With an estimate of the metallicity known, one can adjust the program $(b-y)$ measures for the effects of a metallicity difference relative to the standard relation, with the entire sequence repeated until the adjustments to each drop below some critical limit, which usually happens very quickly. Using a modified version of this approach in Paper I, from 68 F dwarfs it was found that $E(b-y)$ = 0.025 $\pm$ 0.003 (sem). The dominant source of the uncertainty in the quoted accuracy is the existence of two slightly different standard sequences applied to estimate the reddening; the individual sequences supplied $E(b-y)$ with a precision at 0.001 mag, but the absolute values of $E(b-y)$ differ by 0.004 mag. Likewise, the $m_{1}$ and $hk$ indices, dominated by predominantly Fe lines and Ca $H$ and $K$, respectively, produced [Fe/H] = -0.071 $\pm$ 0.014 (sem) and -0.017 $\pm$ 0.008 (sem), respectively. The higher precision for $hk$ is the combined impact of a smaller sensitivity to reddening changes and a higher sensitivity to metallicity changes. 

With the revised membership list and removal of all potential binaries, the sample of F dwarfs drops to 38; after analysis, one additional star (783) with anomalously large [Fe/H] from both $m_1$ and $hk$ was removed from the discussion. (It is intriguing to note that this star, given its magnitude and the high number of observations in each filter, exhibits unexpectedly large scatter among all indices, possibly indicative of either variability or contamination by another star.) Treating the sample in a fashion identical to Paper I generates reddening almost identical to that of Paper I, with $E(b-y)$ = 0.026 $\pm$ 0.004 (sem). The slight increase in the standard error of the mean is dominated by the smaller sample of stars used to construct the average. For metallicity, however, $m_{1}$ and $hk$ generate slightly higher and lower metallicities, respectively, [Fe/H] = -0.053 $\pm$ 0.020 (sem) and -0.023 $\pm$ 0.013 (sem), leading to  weighted average of [Fe/H] = -0.032 $\pm$ 0.015 (sem), identical to the value derived in Paper I. An often forgotten source of uncertainty in the absolute value of this estimate is that the differentials are defined relative to standard relations assumed to have Hyades metallicity and then translated to solar using an adopted Hyades value of [Fe/H] = +0.12. If the adopted scale for the Hyades is different, e.g. +0.15 \citep{CU17}, the cluster values must be adjusted accordingly, i.e. raising the derived NGC 752 value to solar.

\section{M67: Fundamental Properties}
\subsection{Cluster Membership - Astrometry and Radial Velocities}
While there have been numerous astrometric analyses for membership isolation in M67 (see \citet{GE15} for a summary of the multiple ground-based investigations), we will follow a procedure for M67 similar to that for NGC 752, relying on {\it Gaia} as the exclusive astrometric source for isolating cluster members. We begin again with the data sample of \citet{CA18}, cross-matching the 835 original members with the updated DR3 {\it Gaia} parameters. The original cluster averages of 1.137 $\pm$ 0.060 mas, 
-10.983 $\pm$ 0.238 mas-yr$^{-1}$, and -2.958 $\pm$ 0.239 mas-yr$^{-1}$ become 1.153 $\pm$ 0.056 mas, -10.970 $\pm$ 0.213 mas-yr$^{-1}$, and -2.916 $\pm$ 0.226 mas-yr$^{-1}$ for the $\pi$, $\mu$$_{\alpha}$, and $\mu$$_{\delta}$, respectively. Using the stars of Table 4 matched to the DR3 database, potential 
members were selected if their $\mu$$_{\alpha}$ ($\mu$$_{\delta}$) was within 3$\sigma$ of the cluster mean at 0.636 mas-yr$^{-1}$ (0.678 mas-yr$^{-1}$). Stars outside these boundaries were then checked and retained if they were within 3$\sigma$ of the boundary, where $\sigma$ here refers to the individual quoted uncertainty in either the $\mu$$_{\alpha}$ or $\mu$$_{\delta}$. This select sample was then reduced to those stars with $\pi$ within 3$\sigma$ (0.168 mas) of the cluster mean, if $\pi$/$\sigma_{\pi}$ was 10 or higher. Finally, stars were retained if they were within 3$\sigma_{\pi}$ of the parallax boundary. The final astrometric sample is composed of 897 stars brighter than $V$ = 19.2.

As with the discussion of NGC 752 in Sec. 4.1, the next phase of membership restriction is built upon radial velocities to eliminate nonmembers and isolate possible binaries. Unlike NGC 752, however, one has access to the exquisite 40-year comprehensive, high-precision radial-velocity survey of the cluster field to $V$ = 16.5 to a radius of 30\arcmin\ by \citet{GE15}. As noted earlier, \citet{GE15} based their final membership classification upon both the radial-velocity data and the compiled ground-based astrometric surveys available at the time. Since the latter estimates are now superceded by the {\it Gaia} results, we will appeal to \citet{GE15} only for radial-velocity membership probabilities and binarity. A coordinate match with the 897 astrometric members and the radial-velocity catalog produced an overlap of 652 stars to $V$ = 16.5. Of these, 143 were set aside as binary or triple systems. Of the remaining 509, 42 had indeterminate radial-velocity membership probability while an additional 10 had probabilities in single digits, leaving 457 stars as the probable single-star member database.

\subsection{Reddening and Metallicity}
The technique for reddening and metallicity estimation in M67 is the same iterative procedure as that for NGC 752 in Sec. 4.2 and in Paper I, except the photometry for NGC 752 becomes the defining standard, i.e. the goal is a direct determination of $E(b-y)$ and [Fe/H] for M67 relative to NGC 752. This does not eliminate the need to approach the final estimates in an iterative fashion. Comparison of the M67 $(b-y)$ - H$\beta$ data to that of NGC 752 can reveal an offset due to a difference in cluster reddening and/or a difference in metallicity. At the same reddening, a more metal-rich star at a given H$\beta$ will have a redder $(b-y)$. Likewise, $m_1$ and, to a lesser degree, $hk$ at a given H$\beta$ are affected by reddening. A differential reddening between the two clusters must be accounted for before the final metallicity is determined.

As a starting point, the preliminary reddening difference is obtained by comparing the M67 $(b-y)$ - H$\beta$ photometry between H$\beta$ = 2.55 and 2.68 to the mean relation defined by the single-star members of NGC 752 as compiled in Sec. 4.1. From 339 single-star members of M67, the mean difference in the sense (M67 - NGC 752) is 
-0.002 $\pm$ 0.001 (sem) mag in $(b-y)$, implying that if M67 and NGC 752 have the same metallicity, the reddening in the latter cluster is larger in $E(b-y)$ by 0.002 mag. If 12 stars with larger than typical $\delta$$E(b-y)$ are eliminated, the mean offset remains the same, but the uncertainty reduces to $\pm$0.010 mag for a single star.

Turning to the metallicity, the M67 $m_1$(H$\beta$) and $hk$(H$\beta$) data were adjusted for the preliminary difference in reddening between M67 and NGC 752 and then compared to the mean relations as defined by NGC 752, rather than the Hyades, between H$\beta$ = 2.68 and 2.58. The 
low cutoff for H$\beta$, the same cutoff used in deriving the absolute abundances of NGC 752 in Sec. 4.2, is bluer compared to that for the reddening analysis because the  standard relations for both metallicity indicators steepen significantly among the G dwarfs, leading to a large uncertainty in [Fe/H] for a small uncertainty in H$\beta$. By contrast, the $(b-y)$ - H$\beta$ relation for dwarfs remains approximately linear to at least H$\beta$ = 2.55 (see  Figure 8 of Paper 1). 

With the standard relation now defined by the single, main sequence stars of NGC 752 rather than the Hyades, from 256 single-star members, the average difference in [Fe/H] based upon $m_1$, in the sense (M67 - NGC 752), is +0.115 $\pm$ 0.007 (sem) dex. If 7 stars with signifiacntly larger than average residuals are eliminated, the difference becomes +0.106 $\pm$ 0.006 (sem) dex. The comparable numbers for [Fe/H] based upon $hk$ are +0.044 $\pm$ 0.006 (sem) from 256 stars and, with the 7 stars with the larger than average residuals eliminated, $\delta$[Fe/H] = +0.036 $\pm$ 0.005 (sem). Adopting the average offset of $\delta$[Fe/H] = 0.07 dex as the relative abundance of M67 to NGC 752, one can now recalculate the relative cluster reddening. As expected, since a portion of the redder $(b-y)$ values in M67 is attributable to a higher metallicity, the revised reddening differential, in the sense (M67 - NGC 752) becomes -0.005 mag. Applying an  effect equivalent to the new reddening offset to the $m_1$ and $hk$ values of M67 to place this cluster at the same reddening value as NGC 752 reduces the $m_1$ and $hk$ indices of M67, i.e. makes the stars more metal-poor. The revised metallicity differentials from 249 stars now become  +0.091 $\pm$ 0.006 (sem) dex for $m_1$ and $\delta$[Fe/H] = +0.033 $\pm$ 0.006 (sem) dex for $hk$. The revisions are small enough that further iterations become unnecessary. The implication is that M67 is clearly more metal-rich than NGC 752 by $\delta$[Fe/H] = 0.062 $\pm$ 0.006 (sem) dex. Adopting the combined absolute abundance of [Fe/H] = -0.032 $\pm$ 0.015 (sem) derived in Sec. 4.2 for NGC 752 leads to an absolute [Fe/H] = +0.030 $\pm$ 0.016 (sem) dex for M67, again on a scale where the Hyades is at [Fe/H] = +0.12. The final reddening for M67 becomes $E(b-y)$ = 0.021 $\pm$ 0.004 (sem). Note that the dominant source of the uncertainty in the absolute reddening and metallicity for M67 is the baseline uncertainty in the estimates for NGC 752 which are built upon a significantly smaller sample of stars than M67. 

In an absolute sense, the largest uncertainty in the metallicity estimates for M67 and NGC 752 may lie with the absolute [Fe/H] for the Hyades, typically adopted as +0.12 for the photometric calibrations but more recently derived from some spectroscopic analyses as +0.15 \citep{CU17}. Beyond this issue, the most probable source of error lies with the photometric zero-points. The sequence of analyses \citep{JO95, JO97, TA07, TA08} linking the exceptionally accurate Str\"omgren photometry of the Hyades, NGC 752, and M67 discussed in Section 3 should generate zero-point uncertainties in the relative cluster indices as close to 0.000 as possible, with most of the offsets found relative to the older published data as derived in Section 3 being a byproduct of different filters, standards selection, and reduction procedures, a not uncommon issue with all-sky photometry, even when care is taken to minimize such offsets (see, e.g. the Appendix of NTC). A zero-point uncertainty in the $m_{1}$ photometry at the level of $\pm$0.002 mag propagates into an error of $\sim$$\pm$0.022 in the absolute value of [Fe/H]. For $hk$, M67 stars were observed as program stars within the observations used to define the standard system \citep{TA95}, i.e. the night-to-night photometry was transferred to a common system defined collectively by all the stars in the catalog and merged. Since the individual M67 stars were observed over multiple nights and multiple runs, the most likely source of uncertainty in the zero-point arises from the reduced number of observations at fainter magnitudes. To attain the same accuracy in [Fe/H] from $hk$ as defined by $m_{1}$, the uncertainty in the $hk$ zero-point would need to be at least $\pm$0.006 mag. When coupled with the fact that $m_{1}$ is twice as sensitive to reddening effects as $hk$, it seems highly likely that the [Fe/H] from $hk$ is at least as accurate from a zero-point standpoint as that of $m_{1}$. A final point we will return to below is the obvious issue that $m_{1}$ and $hk$ also measure different indicators of metallicity, the former dominated by weak Fe lines and the latter indicative of Ca. 

\section{Cluster Distance and Age}
Before discussing the derivation of the individual cluster distances and ages through parallax and comparison to appropriate isochrones, we first revisit the question of binarity among the cluster members. As noted earlier, the binary evaluation of M67 is unique due to the unusually high precision radial-velocity coverage of virtually all the stars brighter than $V$ = 16.5 for over 40 years \citep{GE15, GE21}. While the sample is less complete, the velocities less precise, and, with the exception of \citet{ME98, ME08, ME09}, the temporal coverage generally random for NGC 752, compared to the vast majority of open clusters, the determination of radial-velocity binarity for NGC 752, as with the Hyades, remains exceptional, in large part due to the proximity of the cluster. For clusters without such detailed radial-velocity insight, the $uvby$H$\beta$ system has long offered an option for identifying potential binary systems composed of stars of comparable mass on the main sequence. For stars significantly redward of a cluster turnoff, such systems are readily identifiable due to their position above the unevolved main sequence to the limit of 0.75 mag. However, as has been illustrated in innumerable cluster CMD discussions, the cluster binary sequence eventually crosses and merges with the vertical turnoff, making it impossible to visually distinguish between a single star evolving off the main sequence or a composite composed of two stars still sitting near the base of the turnoff. This confusion becomes problematic for delineation of the evolutionary path of stars passing through the hydrogen-exhaustion phase (HEP) and post-HEP phase {\it en route} to the subgiant branch since these phases are rapid and few stars are likely to tbe seen during them. Thus, the contamination of this portion of the CMD by a handful of binaries can easily distort the perceived location of the turnoff and the luminosity of the stars populating the blue edge of the subgiant branch. 

The fundamental technique is straightforward. For stars on the unevolved main sequence (ZAMS) at a given metallicity there is a well-defined relation between $(b-y)$ and $c_1$. Since $c_1$ for late A through early G stars is a surface gravity/luminosity indicator, as a star evolves away from the main sequence and its luminosity grows, $c_1$ grows accordingly. Thus, at a given $(b-y)$ for the turnoff region of a cluster CMD, the brighter stars should exhibit a correlation between increasing $c_1$ and decreasing $M_{V}$. Using a mixture of field stars and nearby clusters, including NGC 752, \citet{CR75} initially derived the slope of the relation between $\delta$$c_1$ =  ($c_{1OBS}$ - $c_{1ZAMS}$) and $\delta$$V$ = ($V_{ZAMS}$ - $V_{OBS}$). This slope was revised by NTC using the extensive M67 turnoff among cool F dwarfs and applied to identify unknown binaries in M67 and again in NGC 752 \citep{DA94} using the photoelectric photometry of \citet{TW83}. While the luminosity correction to $c_1$ was crucial for the estimation of cluster distances from $uvby$ photometry of the turnoff stars, with parallaxes available one can focus exclusively on the identification of undetected binaries. Because our concern is solely with the differential comparison of stars within the cluster, the analysis is totally independent of the cluster reddening and metallicity and limited only by the precision of the photometry.

\subsection{NGC 752}
For the first case, we return to the 126 astrometric members of NGC 752. Eliminating the 6 probable radial-velocity nonmembers from \citet{AG18} and all red giants, we are left with 67 stars bluer than $(b-y)$ = 0.55 to $V$ = 14.5, i.e. to mid-G stars. A linear fit was drawn between $V$ and $(b-y)$ for stars fainter than $V$ = 11.2 to the limit of 14.5 to define $V_{ZAMS}$$(b-y)$; the same stars were then used to define $c_{1ZAMS}$$(b-y)$. For all 67 stars, $\delta$$c_1$ and $\delta$$V$ were derived; the results are plotted in  Figure 2. The trend of increasing differential luminosity with increasing differential $c_1$ is obvious, though there are 11 stars (red points) that lie systematically above the predominant trend. In Figure 3 we plot the CMD for the 67 member single and binary stars included in the discussion, tagging the 11 deviants of Figure  2 again as red open circles. Of the 11 discrepant stars, 9 are known binaries. The remaining two, PL 648 and 1003, have not exhibited radial-velocity variability but clearly sit significantly above the main sequence at $V$ = 12.06 and 11.17, respectively. (Star 1003 was first tagged as a probable binary via photoelectric $uvby$ photometry using the same approach, as demonstrated in  Figure 1 of \citet{DA94}. Of the 7 stars tagged as potential binaries in that analysis, with the exception of 1003, all are now known to be binaries and/or nonmembers.) Given the high probability of cluster membership for both 648 and 1003, it is possible that the two systems are inclined at an angle that minimizes the radial-velocity variations of the stars and/or that the orbital period of the system is large.
The value of identifying probable binaries in NGC 752 is best illustrated by the stars at $V$ = 11 and brighter. For the red points with $(b-y)$ \textgreater\ 0.30 and 
fainter than $V$ = 11 in  Figure 3, any precision CMD would reveal their probable binary nature given their position significantly above the ZAMS. For the five red points at the top of the turnoff, their removal narrows the spread in $V$ among the stars defining the red hook at the turnoff and eliminates the possibility of constraining the post-HEP and subgiant branch beyond this point using star 1117 at $V$ = 9.6, a known SB2 system. 

\begin{figure}
\figurenum{2}
\includegraphics[width=\linewidth,angle=0]{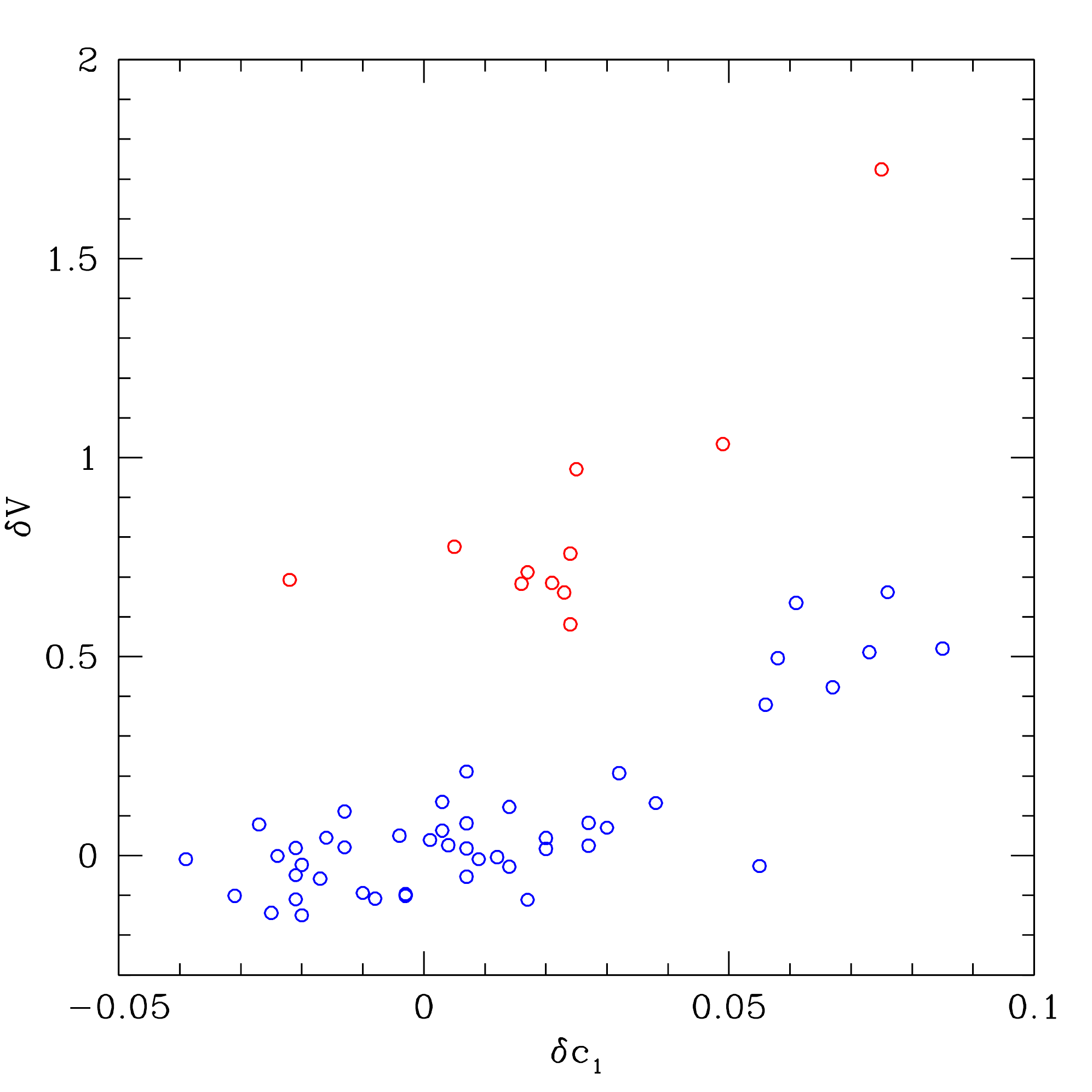}
\caption{Correlation between distance in $V$ above the ZAMS with the change in $c_1$ for stars at the turnoff in NGC 752. Blue points define the relation for single stars. Red 
points identify probable binaries.}
\end{figure}
\begin{figure}
\figurenum{3}
\includegraphics[angle=0,width=\linewidth]{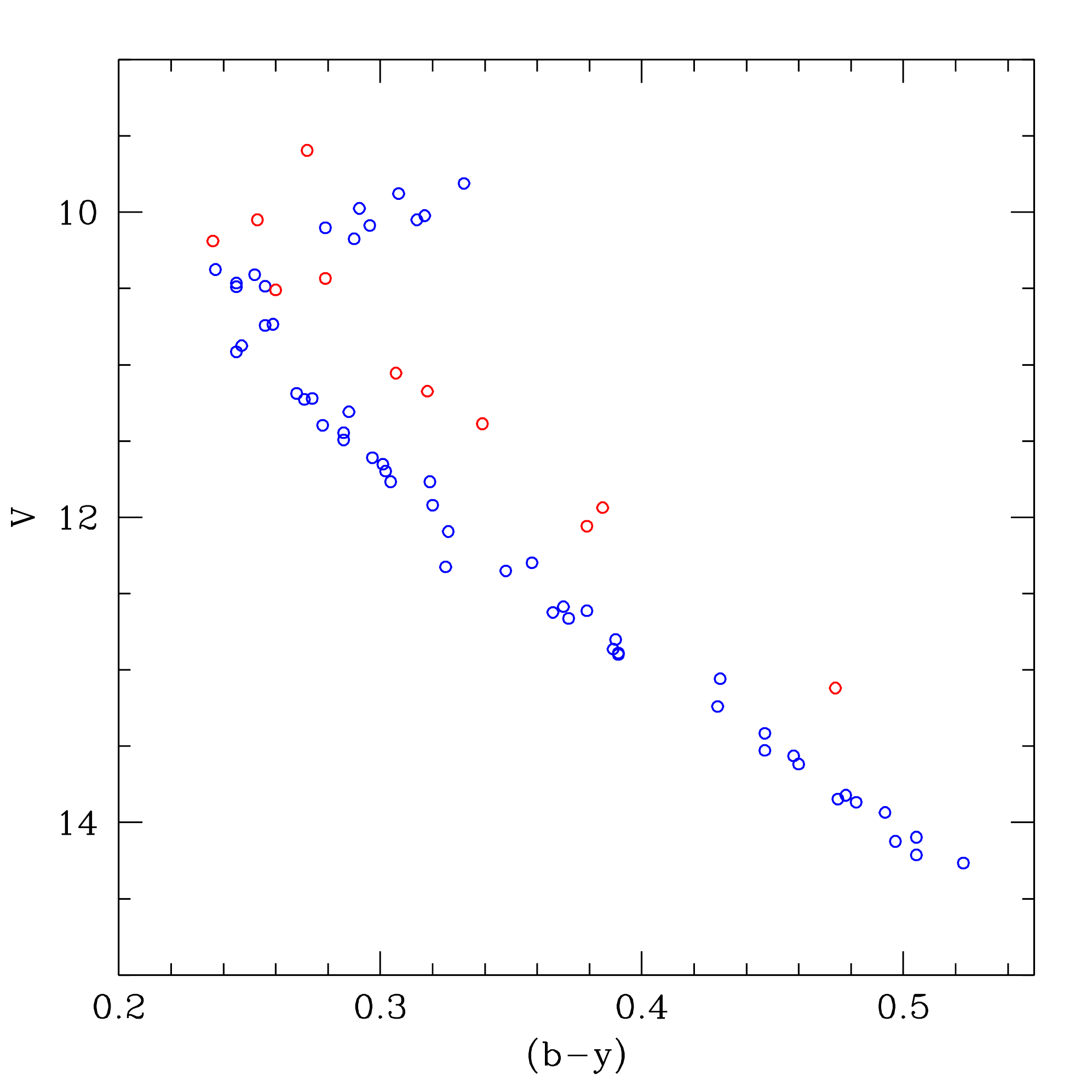}
\caption{CMD for stars in Figure 2. Symbols have the same meaning.}
\end{figure}

To close, we determine the age of the cluster using the CMD with all nonmembers and/or binaries, photometric or spectroscopic, eliminated. For isochrones, we adopt the same set \citep{VA06} discussed in Paper I, interpolated slightly between [Fe/H] = 0.000 and -0.039 to match the derived value of [Fe/H] = -0.032. For the distance, use is made solely of the {\it Gaia} DR3 parallax data. If the parallaxes for the 120 members discussed above are averaged, the mean $\pi$ is 2.260 $\pm$ 0.076 (sd) mas or $(m-M)_{0}$ = 8.23. Since $\sigma_{\pi}$ grows larger on average with increasing $V$, we can first restrict the sample to members brighter than $V$ = 16.0, generating an average $\pi$ = 
2.267 $\pm$ 0.065 (sd) mas or $(m-M)_{0}$ = 8.22 from 96 stars. Finally, one can limit the sample to only stars where $\pi$/$\sigma_{\pi}$ \textgreater\ 50. From  83 stars, the average $\pi$ = 2.270 $\pm$ 0.046 (sd) or $(m-M)_{0}$ = 8.22 $\pm$ 0.04 (sd). With $E(b-y)$ = 0.026, the apparent modulus becomes $(m-M)$ = 8.33 $\pm$ 0.04 (sd). It is important to recognize the increase in the parallax for NGC 752 (0.047 mas) relative to the mean value for the stars from \citet{CA18}. Systematic offsets at this level have been applied to a number of  clusters in this series analyzed using \citet{GA18} (DR2) data, NGC 6819 \citep{DE19}, NGC 7142 and M67 \citep{SN20} (Paper II), and NGC 2243 \citep{AT21}, with the common justification for these adjustments supplied by \citet{RI18, ST18, ZI19}, among others.

Figure 4 shows the resulting CMD-isochrone comparison covering the full range of the CMD. The fit to the isochrones is excellent over the $M_{V}$ range from the giant branch to $M_{V}$ $\sim$ 5. Toward fainter magnitudes, the observed points lie above the models by an amount that increases toward fainter magnitudes. Such discrepancies are common for cooler dwarfs in comparisons with theoretical isochrones due to the difficulty of transferring cool star models to the observational plane because of the complexity of cool dwarf atmospheres, challenging bolometric corrections, and the construction of synthetic color indices on intermediate and narrow-band systems. A similar trend is seen in a comparison of the earlier {\it Gaia} photometry of  NGC 752 using the same models adopted here but transferred to the {\it Gaia} photometric system (see Figure 9 of \citet{BO20}). An empirically corrected ZAMS for the isochrones of the cooler dwarfs in NGC 752 has been derived and plotted in Figure 4 as a green dashed curve. Such an approach has long proven valuable in adjusting theoretical isochrones to more realistically match observed stellar photometry in temperature and luminosity space where theoretical transformation relations may be inadequate (see, e.g. \citet{PI04, AN07}).  We will return to this issue in Sec. 6.2.

\begin{figure}
\figurenum{4}
\includegraphics[angle=0,width=\linewidth]{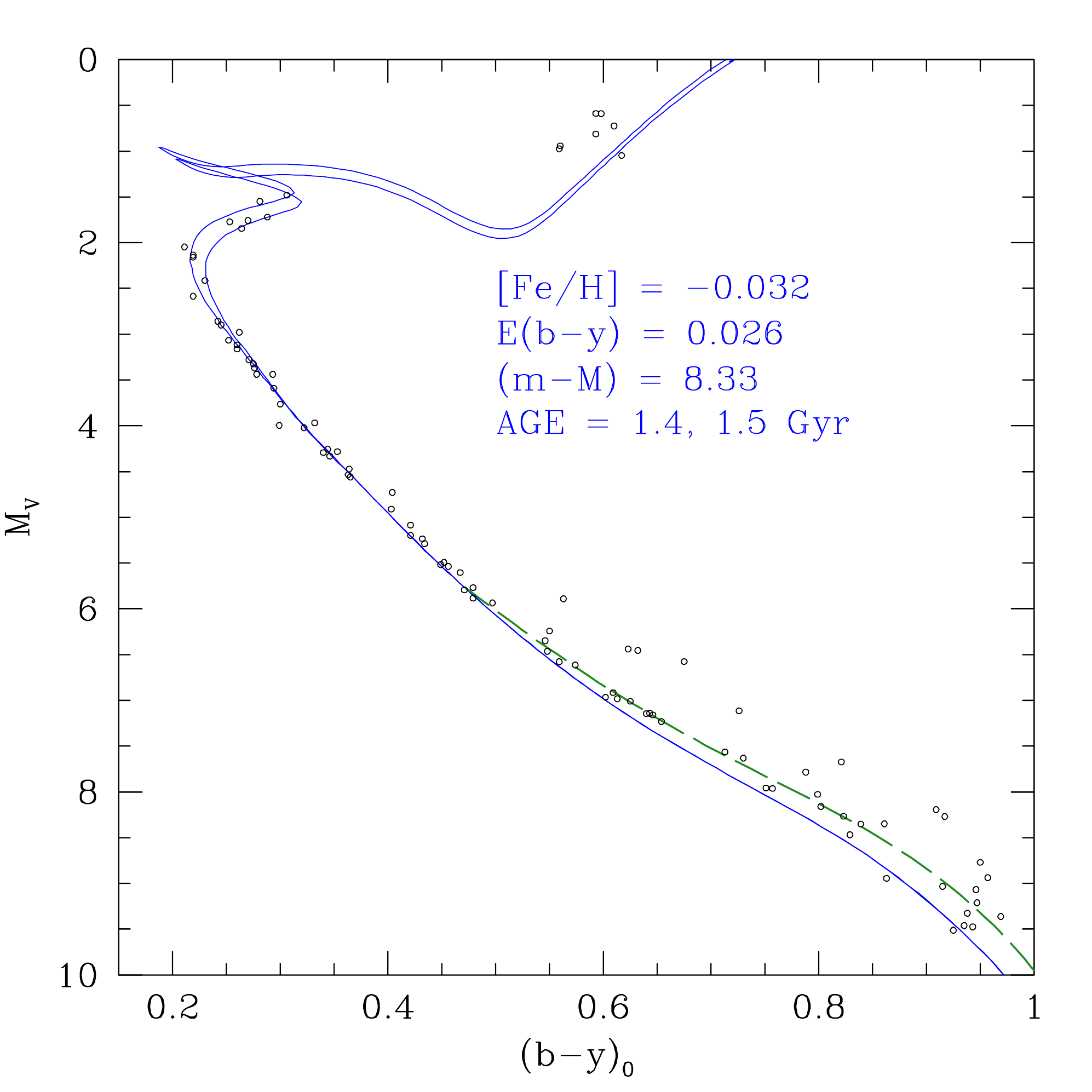}
\caption{Age derivation for NGC 752 from single stars using $(m-M)_{0}$ derived solely from parallax, coupled with photometric reddening. Isochrones are 
metallicity adjusted models of VR. Dashed green line represents an empirically derived correction to the isochrone lower main sequence.}
\end{figure}

As expected,  quality of the fit between the observations and theory in Figure 4 is identical to that in the pre-{\it Gaia} Paper I analysis, though the clarity of the CMD is enhanced by the removal of binaries and probable nonmembers. Since the best-fit $(m-M)$ of 8.30 $\pm$ 0.05 with $E(b-y)$ = 0.025 led to a tight age range of 1.4 to 1.5 Gyr in Paper I, the minor alterations to the fundamental cluster parameters have little impact on the current derivation. 

For comparison, \citet{AG18} used MINESweeper, a Bayesian approach for determining stellar parameters with MESA Isochrones \& Stellar Tracks (MIST) evolutionary models \citep{CH16, DO16} to infer probability distribution functions for the age and distance of each of 53 single cluster members in NGC 752. MINESweeper provided full posterior distributions of all predicted stellar parameters from the MIST models, including ages, masses, and radii, leading to 
$(m-M)_0 = 8.21^{0.04}_{0.03}$, [Fe/H] = $+0.02 \pm 0.02$, $A_{V} = 0.198^{0.008}_{0.009}$, and an age of 1.34 $\pm$ 0.06 Gyr. The true distance modulus, metallicity, and age all overlap at the $\pm 1\sigma$ level. 
However, the derived $A_{V}$ implies a reddening that is 8$\sigma$ larger than derived in Paper I. Differences in adopted isochrones and photometric systems aside, the excessive reddening derived for the average star in the sample is consistent with the younger age at higher metallicity for NGC 752, though the adoption of too low an overshooting parameter may be the dominant factor, as discussed by \citet{BO20}.

A more recent analysis of the cluster age that made use of the {\it Gaia} DR2 astrometry to identify cluster members is that of \citet{BC20}. Of equal importance is the derivation of the cluster metallicity from high-resolution (R $\sim$ 45000) near-IR spectra of 10 red giants. Adopting the derived reddening of Paper I, the mean [Fe/H] values from optical Fe I and Fe II lines are +0.01 $\pm$ 0.07 (sd) and -0.06 $\pm$ 0.04 (sd), respectively, while the IR Fe II lines generate 0.00 $\pm$ 0.06 (sd). It should be noted that the [$\alpha$/Fe] abundances for light elements like Ca, the source of the $hk$ index, are above solar, typically +0.06 to +0.10 dex. This may be an indication that the higher metallicity from $hk$ relative to $m_1$ is tied to a real metallicity offset for the two indices, but the differential is small enough that it falls within the combined uncertainty of the photometry and the spectroscopy. 

Adopting solar metallicity, a cluster age of 1.52 Gyr was obtained from isochrone fits to  Victoria-Regina \citep{VA06, VA14} (VR) isochrones and MESA \citep{PA11, PA13} models on the {\it Gaia} $G$, $(G-G_{RP})$ photometric system. The true distance modulus adopted by \citet{BC20} for the fit was $(m-M)_0$ = 8.26, derived from the parallaxes of stellar members of NGC 752 in DR2, as discussed in Sec. 4.1. Given the differences in the photometric systems and the changes in the adopted isochrones, the agreement is excellent.

In a contemporaneous discussion of the NGC 752 age via CMD fits to isochrones as part of the interpretation of the masses estimated from detached eclipsing binaries, \citet{SA23} derive a preferred age near 1.6 Gyr using PARSEC isochrones \citep{BR12}, with an uncertainty of about 0.08 Gyr based upon the scatter among the stars in the brightest region of the main sequence, assuming solar metallicity and $E(B-V)$ = 0.044. Since the PARSEC models have similar amounts of overshooting to the Victoria-Regina set, this seems an unlikely source for the difference in age. The simplest solution for the discrepancy is that suggested by \citet{SA23}, the adopted abundance for solar metallicity: the PARSEC models assume $Z_{\sun}$ = 0.0152 while VR models imply 0.0188. The most recent reexamination of the solar value by \citet{MA22} implies 0.0177. If correct, the isochrones used in Figure 4 are, fortuitously, almost identical to the revised solar value ([Fe/H] = -0.006), while the PARSEC isochrones have [Fe/H] = -0.066, leading to an older age at a given turnoff color.

Finally, \citet{LU19} derived abundances for 23 dwarfs and 6 red giants in NGC 752 from high-resolution (R$\sim$ 48000) HIRES spectra, finding [Fe/H] = -0.01 $\pm$ 0.06 (sd) 
and no difference between the giants and the dwarfs. As in \citet{BC20}, the 6 red giants show some enhancement of Ca relative to Fe with [Ca/Fe] = +0.05 $\pm$ 0.03 (sd). 
However, the combined sample of 29 dwarfs and giants produces [Ca/Fe] = +0.02 $\pm$ 0.04 (sd), implying a solar ratio within the errors.

\subsection{M67}
To identify possible undetected binaries near the turnoff of M67, we begin with the sample of 457 single-star members isolated in Sec. 5.1. To ensure as pure a sample as possible, we raise the radial-velocity membership limit to 50\% and remove all stars classed as photometric variables, blue stragglers, and/or X-ray sources, reducing the sample to 421 stars. Since our primary interest is in the stars that populate the vertical turnoff, we finally restrict the photometry to stars between $(b-y)$ = 0.32 and 0.46 with $V$ \textgreater\ 12.5, leaving 253 stars. To ensure that any identifiable CMD features 
are unlikely to be caused by photometric scatter, we have also tested our $V$, $(b-y)$ photometry against the $G$, $(B_{P}-R_{P})$ data of {\it Gaia} DR3. Due to the modest range in color and luminosity under discussion, transformation between the two systems should be possible using little more than a linear color term and an offset \citep{AT21}. The transformation relation from $G$ to $V$ with 1 star excluded exhibits a scatter of $\pm$0.007 mag from 252 stars. The analogous relation for $(B_{P}-R_{P})$ to $(b-y)$, with 2 stars excluded, has a scatter of $\pm$0.004 mag. We have converted the {\it Gaia} photometry to the $V$, $(b-y)$ system and averaged them with the observed $V$, $(b-y)$ data. The two stars exhibiting discrepancies between the two samples have $y$ and $b$ observations numbering in single digits, a possible indicator of potential issues with the photometry given the brightness of these stars. 

As with NGC 752, the CMD of the 251 turnoff stars was used to define $V_{ZAMS}$ between $(b-y)$ = 0.32 and 0.46, and the ZAMS sample then adopted to define $c_{1ZAMS}$ over the same color range. For each star, $\delta$$V$ and $\delta$$c_{1}$ were derived; the result is plotted in Figure 5, where the blue and red points are from the averaged, single-star data. Somewhat surprisingly, despite the comparable precision of the photometry in M67 relative to NGC 752 for the same class of stars located 1.5 mag fainter in the former cluster, the separation into single and possible binary stars is less obvious. Moreover, the $\delta$$V$ - $\delta$$c_{1}$ distribution in Figure 5 appears different from that in Figure 2. The two primary contributors to this are the presence of stars defining the blue hook and a subgiant branch extending to $(b-y)$ = 0.46, neither of which are included in the NGC 752 CMD, and the declining slope of the $\delta$$V$/$\delta$$c_{1}$ relation with increasing $(b-y)$ (NTC). The former phenomenon ensures that a significant fraction of the stars redder than the turnoff sit at increasingly larger distance above the ZAMS with increasing $(b-y)$. In addition to the basic change in the slope of the relation with increasing color, because the turnoff of M67 is significantly cooler than that of NGC 752, the latter phenomenon ensures that a comparable error in $c_{1}$ generates a larger spread in $\delta$$V$. To illustrate the continued value of the data presented in Figure 5, the photometry for all systems classified as SB2 or triplet near the turnoff of M67 by \citet{GE15} has been processed in the same manner as the single stars, leading to the green symbols in Figure 5. The separation of these points from the band defined by the majority of the single stars is obvious. Using the green symbols as the defining binary sample, stars that lie above the approximate boundary between the green and blue points have been characterized as probable photometric binaries, with the likelihood of a correct identification increasing with increasing distance above the blue band.

\begin{figure}
\figurenum{5}
\includegraphics[angle=0,width=\linewidth]{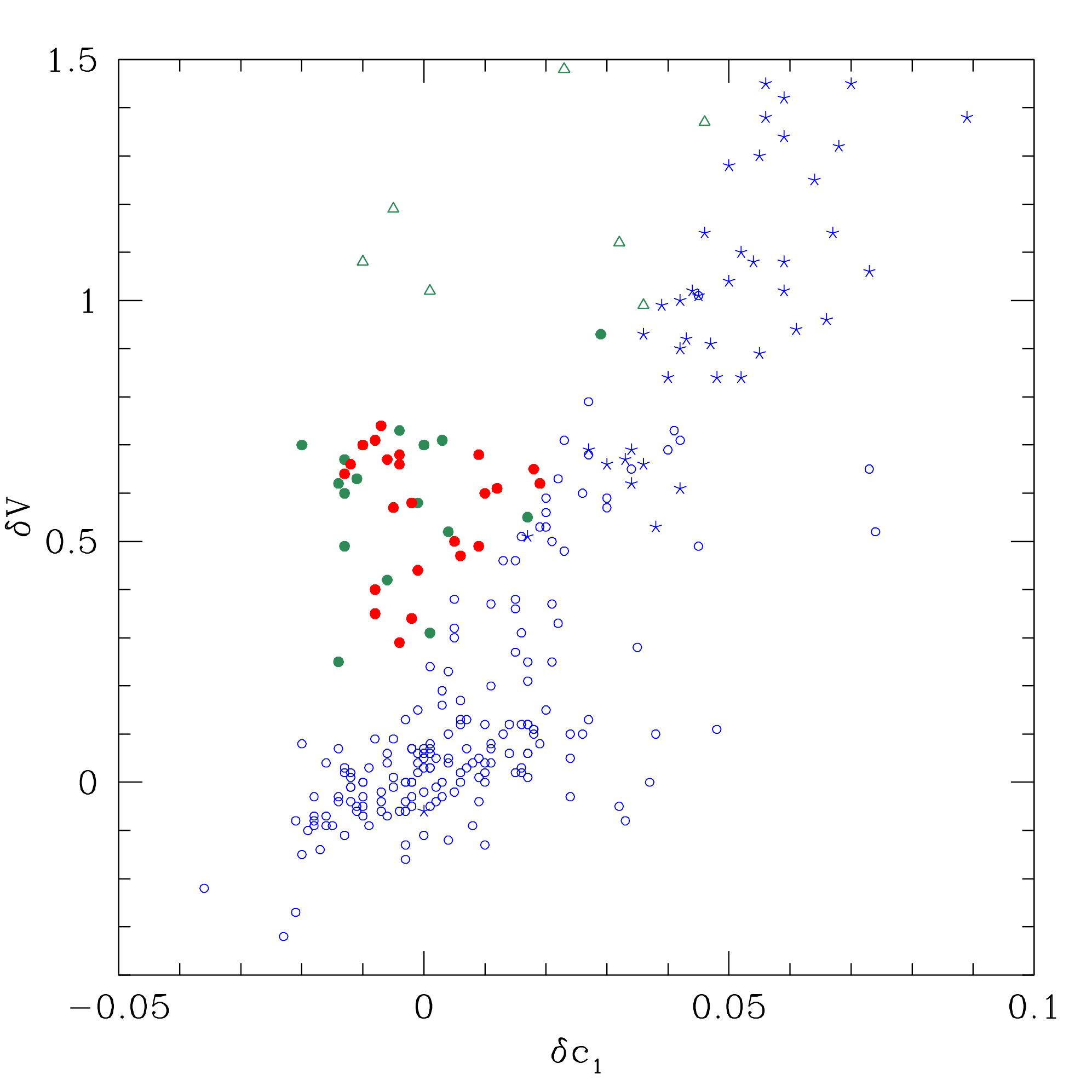}
\caption{Same as Figure 2 for M67. Blue and red symbols show stars classed as single and probable binary stars, respectively. Green symbols show stars classed via radial 
velocities as SB2 or triples. Blue asterisks show the location for single subgiants, while green open triangles indicate binary subgiants.}
\end{figure}

\begin{figure}
\figurenum{6}
\includegraphics[angle=0,width=\linewidth]{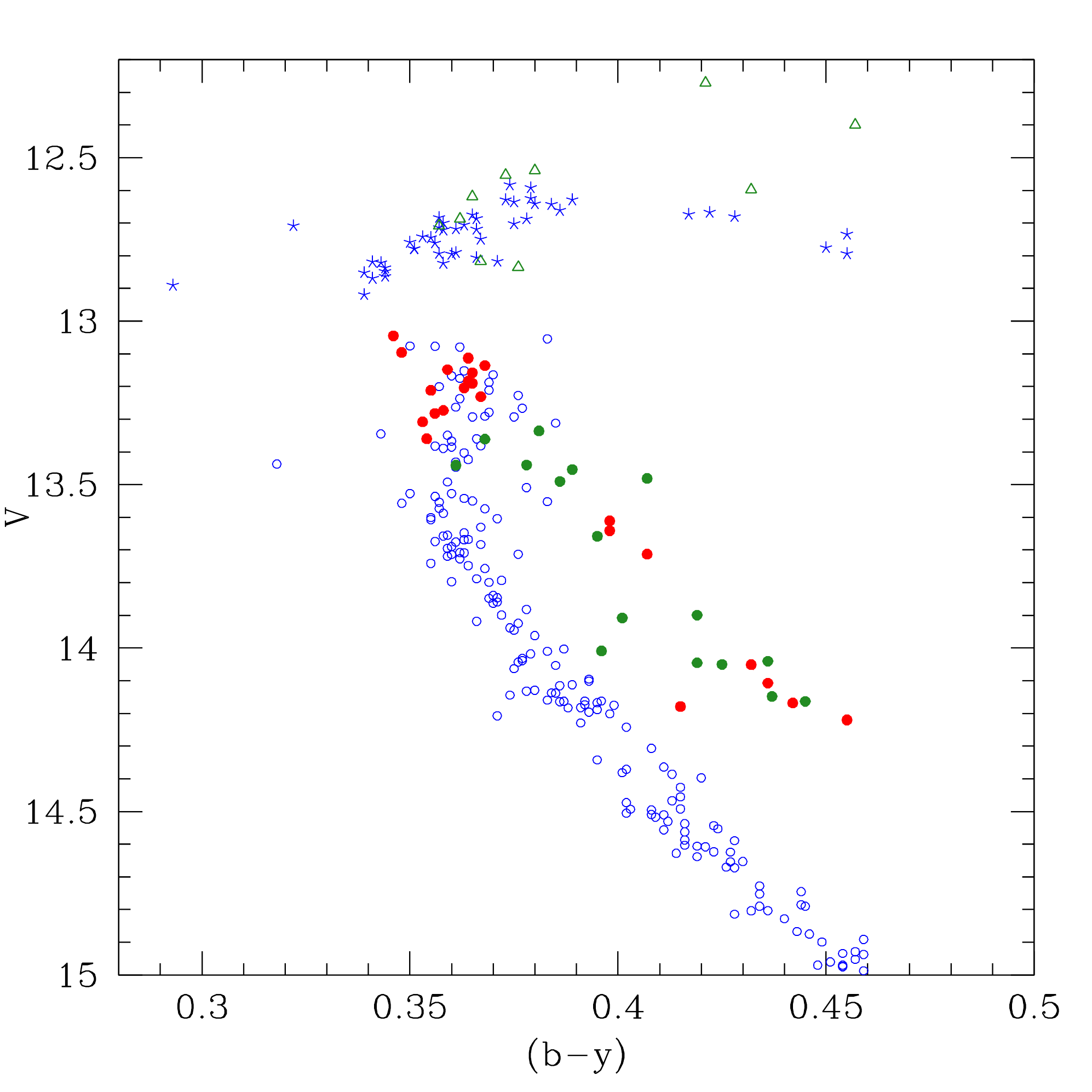}
\caption{Same as Figure 3 for M67. Symbols have the same meaning as in Figure 5.}
\end{figure}

In Figure 6, the CMD for the stars of Figure 5 is presented; the symbols have the same meaning including the use of blue five-pointed stars for single subgiants and green open triangles for likely binaries. As in Figure 3, the combination of photometry and spectroscopy eliminates the majority of the well-defined sequence of isolated stars sitting sigificantly above the ZAMS, though it should be remembered that, based upon radial velocity, all the red points above the ZAMS are classed as single stars. The second largest concentration of possible binaries lies, as one might expect, along the binary sequence extension into the main sequence hook. It should be emphasized that these stars fall in the zone just above the blue band of Figure 5, making their binary nature less probable than the obvious band of stars sitting 0.5 to 0.7 mag above the single-star relation in Figure 6. We close this discussion by noting that there are a number of stars that lie in apparently anomalous locations of the CMD without any photometric or spectroscopic indications of binarity, an issue returned to below.

As with NGC 752, the age of the cluster is best determined from the CMD after removal of all potential binaries or triples, photometric or spectroscopic, variables, blue stragglers, X-ray sources, and other known anomalies. For the cluster distance we again rely solely on the {\it Gaia} DR3 parallax data. As discussed in Sec. 5.1, the full set of astrometric members has an average $\pi$ = 1.153 $\pm$ 0.056 (sd) mas. If only the stars brighter than $V$ = 16 are included, the mean becomes 1.154 $\pm$ 0.052 (sd) mas. Finally, if only 312 stars with $\pi$/$\sigma_{\pi}$ \textgreater\ 50 are used, $\pi$ = 1.162 $\pm$ 0.032 (sd) mas. With $E(b-y)$ = 0.021, the latter two determinations lead to $(m-M)$ = 9.76 and 9.78, respectively; $(m-M)$ = 9.77 will be assumed. As with the earlier discussion of NGC 752, the final parallax from DR3 data is larger by 0.025 mas than the estimate from the original \citet{CA18} parallaxes, a smaller offset than for NGC 752, but well within the range for other open clusters. For photometry, the $V,(b-y)$ data of Table 4 will be used except at the turnoff where the averaged results illustrated in Figure 6 will be given precedence. The CMD superposed on isochrones of age 3.5 and 3.7 Gyr adjusted to an assumed [Fe/H] of 0.03 is shown in Figure 7. The isochrones supply an overall excellent match to the observations down to the main sequence near $M_{V}$ $\sim$ 5. For fainter stars, the isochrones are increasingly too faint compared to the data, the same pattern seen for NGC 752. Superposed on the plot (dashed green curve) is the empirically derived ZAMS correction to the isochrones as defined by the lower main sequence data of NGC 752. This correction leads to an excellent match between theory and observation to the limit of the plot.

It should be noted that, unlike the subgiant branch and the unevolved main sequence, the first-ascent red giant branch for the isochrones supplies a less satisfying match to the cluster photometry. Unlike NGC 752, which has no blue stragglers or subgiants and only one probable first-ascent red giant, M67 is rich in blue stragglers and composite systems near the turnoff. The evolved counterparts of these anomalous systems are the likely source of the scatter at the base of the red giant branch between $M_{V}$ = 2 and 3, making the exact location of the single-star red giant branch difficult to define. More important, as with the red dwarfs, conversion of the isochrones from the theoretical to the observational plane requires transformation relations between temperature/luminosity and $b-y$ as a function of metallicity. While these have been well developed for G dwarfs and hotter, as with the red dwarfs, they are less reliably defined for cool giants and likely require an empirical correction of the type developed for the dwarfs to better reproduce the true Str\"omgren system.

\begin{figure}
\figurenum{7}
\includegraphics[angle=0,width=\linewidth]{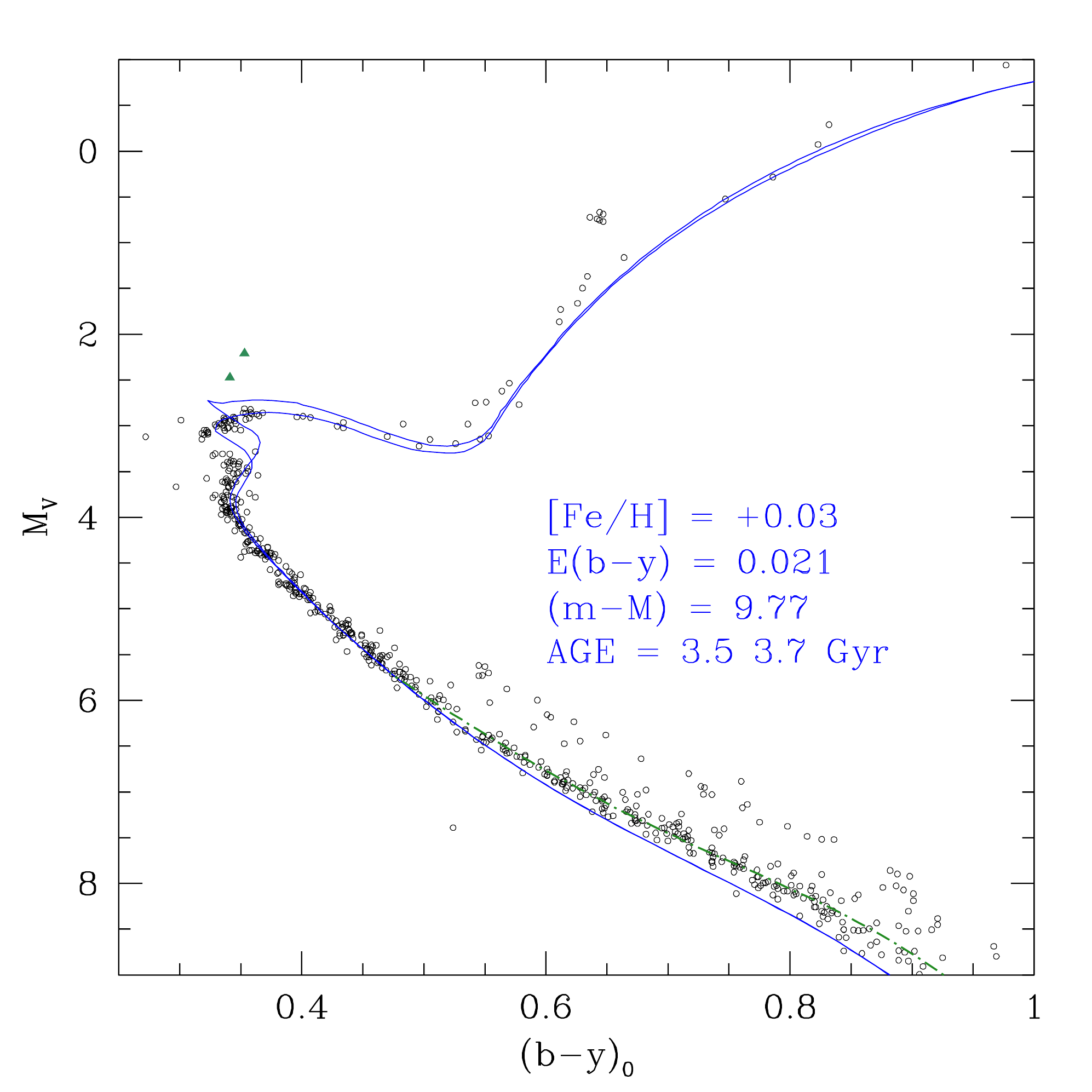}
\caption{Cleaned CMD composed of probable members of M67, adjusted for $(m-M)$ = 9.77 and $E(b-y)$ = 0.021 and compared with isochrones of age 3.5 and 3.7 Gyr. Dashed green curve is the corrected cooler ZAMS of Figure 4 for NGC 752 superposed on the adopted isochrones. Green triangles represent additional photometric binaries.}
\end{figure}

Before moving to the turnoff, three items should be mentioned. First, the lack of binary information for $M_{V}\sim 5.5$ is apparent given the reemergence of the usual photometric binary sequence for the ZAMS below this point. Second, the isolated star below the main sequence near $M_{V}$ $\sim$ 7.3 is an almost certain nonmember. Due to the absence of radial velocities at this apparent magnitude level, membership depends entirely upon astrometry. This star sits at the very boundary of the astrometric limits used to separate members from nonmembers. Third, two stars above the blue end of the subgiant branch are plotted as green triangles, implying probable binaries. These two stars were excluded from the earlier $V$, $c_{1}$ discussion because they were situated well above the subgiant branch and their inclusion extended the $\delta$$V$ scale of Figure 5 by an additional 0.75 mag, giving exaggerated emphasis to the evolved stars. A simple check allows us to determine if these systems are likely composites of a pair of subgiants. From 29 supposedly single subgiants between $(b-y)$ = 0.35 and 0.38, the average $(b-y)$ and $c_{1}$ indices are 0.365 $\pm$ 0.009 (sd) and 0.433 $\pm$ 0.009 (sd), respectively. The two stars represented by the green triangles have ($b-y, c_{1})$ = (0.362, 0.426) and (0.374, 0.443), identical within the uncertainties despite sitting 0.48 mag and 0.67 mag, respectively, above the subgiant branch.

The expanded turnoff region of Figure 7 is presented in Figure 8. Unlike NGC 752, the isochrone profile of the turnoff hook provides a less satisfying match to the data. 
While the magnitude level of the bluest point of the turnoff below the HEP can be considered consistent with either isochrone, the luminosity of
the subgiant branch is almost perfectly matched by the 3.7 Gyr isochrone. The color of the turnoff data places it blueward of either isochrone, implying an age younger than 3.5 Gyr, though such an age would be contradicted by the need for a subgiant branch even brighter than the already too bright 3.5 Gyr isochrone. Thus, the key to defining the cluster age is the distinction between the isochrones of Figure 4 versus those of Figure 8. Keep in mind that for clusters in the 1-2 Gyr range, as in NGC 752, the rapid evolution of turnoff stars after reaching the end of the red hook often leaves the blue hook and the bluer portion of the subgiant branch poorly populated, if at all. Thus, the defining pattern of the younger isochrones with increasing age is a correlated shift to the red and fainter magnitudes. By contrast, the color band defined by the red hook and post-HEP phase among isochrones near 4 Gyr evidences very little color evolution with increasing age.  For solar metallicity models, the range in 
$(b-y)$ between the blue hook and the limit of the red hook goes from $(b-y)_{0}$ = 0.316 to 0.359 for 3.5 Gyr, to 0.320 to 0.353 for 3.7 Gyr, to 0.321 to 0.345 for 3.9 Gyr. By contrast, $M_{V}$ of the subgiant branch at the color of the blue limit below the HEP shifts from 2.75 to 2.90 to 3.03 over the same age range. As clearly demonstrated in Fig. 8, with precision photometry $M_{V}$ changes at this level are easily detectable, particularly once the confusion from binaries and variables is eliminated. On an absolute scale, the availability of {\it Gaia} parallaxes and cluster membership for a stellar sample numbering in the hundreds reduces the uncertainty in $(m-M)_{0}$ to the level of 
$\pm$0.01 mag, independent of both reddening and metallicity. The primary uncertainty in the age estimate tied to the luminosity of the subgiant branch becomes  
the reddening since the observed CMD must be adjusted for extinction. For M67, due to the exceptional accuracy of the reddening estimate, this adds no more than $\pm$0.02 mag to the uncertainty of the position of the subgiant branch. Thus, the estimated age of M67 defined by the luminosity of the subgiant branch and tied to the specific isochrones and cluster parameters of Figure 8, is 3.70 $\pm$ 0.03 Gyr. Until the issue with the color morphology of the turnoff is clarified, a more conservative estimate of the uncertainty based upon the spread in color at the turnoff below the HEP is 3.7 $\pm$ 0.1 Gyr. 

\begin{figure}
\figurenum{8}
\includegraphics[angle=0,width=\linewidth]{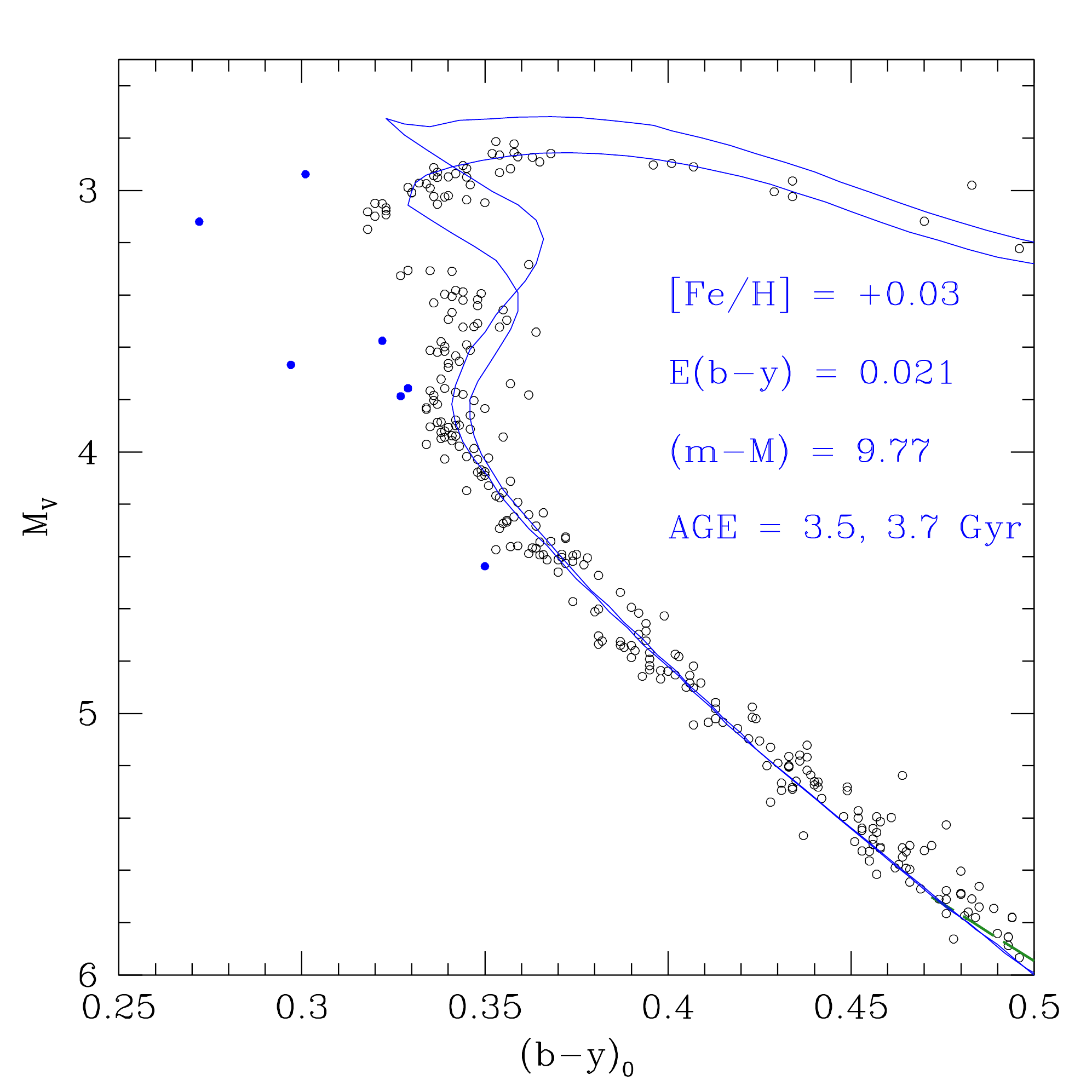}
\caption{Expanded turnoff region of the M67 CMD of Figure 7. Blue points are stars with anomalously blue colors.}
\end{figure}

How do these results compare with other analyses? Rather than generating a litany of derived ages and distances from multiple sources using different stellar models under varying assumptions about the cluster age and reddening, we will focus on two distinctly relevant investigations at opposite ends of the stellar sample scale. The first is Paper II. Paper II derives the fundamental cluster parameters of NGC 7142 using precision multicolor broadband photometry compared to the multicolor data of the Hyades and through a differential CMD comparison to a virtually identical cluster, M67. As part of the analysis, the authors also redetermine the absolute paramters for M67, making use of the Yale-Yonsei isochrones \citep{DE04} in multiple colors to constrain the distance and age. From the multicolor index comparisons to the Hyades with the standard star cluster data of M67 from \citet{ST00}, Paper II finds $E(B-V)$ = 0.04 $\pm$ 0.01 and [Fe/H] = -0.02 $\pm$ 0.05, on a scale where the Hyades has +0.15. The age and distance results are 3.85 $\pm$ 0.17 Gyr and $(m-M)$ = 9.75, respectively. To compare with our age and distance results, the reddening must be lowered to $E(B-V)$ = 0.028 and the metallicity increased by +0.08 dex, keeping in mind that our scale has the Hyades at +0.12. The lower reddening decreases the apparent modulus by 0.07 mag, while boosting the metallicity requires a larger distance by $\sim$0.08 mag \citep{TW09}, changing the final apparent modulus to 9.76. Likewise, lowering the reddening leads to a redder turnoff, but higher metallicity at a given turnoff color generates a lower age (Paper II). While the consistency between the two sets of photometric data from two distinctly different sets of isochrones is encouraging, the more relevant aspect of the analysis is the CMD comparison between the isochrones and the M67 fiducial points (Figure 15 of Paper II). The data for M67 are compared to isochrones of age 3.5 and 4.0 Gyr. What is apparent upon close examination of the turnoff is that while the points below the HEP follow the trend expected for a 3.6 Gyr isochrone, they do not reproduce the blue hook, and populate the subgiant branch at a luminosity consistent with an age of 3.9 Gyr. As is also the case with VR (below), one cannot simultaneously match the color of the turnoff, the shape of the blue hook, and the luminosity of the subgiant branch. Shifts to the red to optimize the color of the turnoff reduce the apparent modulus by an amount that contradicts the limits of the {\it Gaia} parallaxes, while moving the red giant fiducials away from the isochrones, not closer. Note, a similar argument can be made for the fit in Figure 8. A shift to the red for the $(b-y)_{0}$ photometry by reducing $E(b-y)$ to 0.011 mag would provide a reasonable color match between the observations and the 3.7 Gyr isochrone, but the correlated shift downward to realign the ZAMS data with the isochrone makes the observed subgiant branch too faint by more than 0.1 mag compared to the isochrones and again reduces the distance modulus to a value outside the allowed limits of the parallaxes.

To search for an alternative solution to the morphology issue at the turnoff, we turn to the detailed and incisive investigation of a single M67 binary system by SA. (In the discussion that follows, identifications in M67 are given using WOCS numbers from \citet{GE15}.) Star 11028 was selected for analysis by SA as a double-lined spectroscopic binary near the cluster turnoff with a large enough semimajor axis to minimize the likelihood of mass transfer and/or tidal distortions between the two members, i.e. both stars followed the evolutionary path of isolated single stars. From Kepler K2 observations \citep{ST16}, only one eclipse is observed, but the inclination angle is so well constrained that precise mass estimates can be achieved. SA adopted $E(B-V)$ = 0.041 \citep{TA07} and, after discussing the issue of systematic errors in DR2 parallaxes, adopted $(m-M)_{0}$ = 9.63 $\pm$ 0.06. Fortuitously, the difference in $E(B-V)$ compared to our value almost exactly compensates for the lower absolute modulus, leading to $(m-M)$ = 9.76. Using multiwavelength spectral energy distributions (SED), SA identified possible singe-star pairs of stars from the observed M67 CMD which, when combined, best matched the composite 11028 system. The optimal fit for the primary component placed the star at the bluest point of the turnoff below the HEP, a critical location ideally suited to constrain the cluster age because the CMD turnoff at this location is effectively vertical. Given the distance, $T_{\rm{eff}}$ from the SED, the radial-velocity and eclipsing light curves, one can tightly constrain the mass, radius, and luminosity of the primary. Beyond this point, SA encountered for their single pair of stars the problem equivalent to the deviant morphology of the turnoff seen in Figure 8.  Their problem is that the well-constrained primary mass, 1.222 $\pm$ 0.006 $M_{\sun}$, $T_{\rm{eff}}$, and luminosity combine to predict an age for the star (3.0 $\pm$ 0.3 Gyr) that is below the consistently derived 3.5 to 4.5 Gyr range. Because stars of a given mass generally grow more luminous as they evolve off the main sequence and approach the HEP, a younger age implies a lower luminosity, i.e. the primary star is too faint for the cluster age or, equivalently, the radius of the star is too small for its mass. This is exactly the same pattern seen for the stars at the blue limit below the HEP in Figure 8. The primary single-star analog, 6018, sits at ($(b-y)_{0}$, $M_{V}$) = (0.334, 3.837), exactly at the blue limit below the HEP. Alternatively, the mass of a star at this luminosity  from our isochrones ranges from 1.191 $M_{\sun}$ to 1.184 $M_{\sun}$ for isochrones between 3.5 and 3.8 Gyr. While raising the metallicity of M67 moves the mass in the correct direction (SA), even adopting the Hyades metallicity for M67 comes up short with the predicted mass range from 1.211 to 1.208 $M_{\sun}$ over the same age range. The added flaw in this solution is that boosting the metallicity to as high a value as the Hyades forces a 0.1 mag increase in the distance modulus from main sequence fitting to the isochrone ZAMS. With the true distance modulus set by {\it Gaia}, this increase in $(m-M)$ can only be taken up by increasing the reddening. The combined shift to the blue for the observed data places the points systematically below the isochrones. Oddly, the observed data can be forced to match the Hyades isochrone below the level of the HEP if $(m-M)_0$ = 9.68 and reddening is set to 0. Unfortunately, the position of the observed HEP is 0.5 mag too faint and the shape of the HEP phase is distinctly different from the isochrone. It should be noted that that the mass range for stars on the giant branch for the 3.7 Gyr isochrone of Figure 7 is 1.357 to 1.395 $M_{\sun}$; for a Hyades metallicity, the analogous range is 1.385 to 1.42 $M_{\sun}$. At an age near 4 Gyr, \citet{ST16} measured an average asteroseismic mass for the red giants of 1.36 $\pm$ 0.01 $M_{\sun}$.

SA also offer the possibility that the binary discrepancy is due to a significantly lower degree of convective overshoot and/or diffusion. While we can offer no additional insight beyond the excellent discussion of the former option by SA, the latter may have relevance for the primary component. Diffusion can only occur in an atmosphere that lacks significant levels of mixing triggered by convection and/or rotation. As regularly noted in previous papers of this series designed to probe the Li evolution of open clusters, standard stellar evolution theory predicts only a small amount of convection for stars at the mass of the M67 turnoff and higher, but increasing levels of mixing for stars of 1.2 $M_{\sun}$ and lower \citep{DE90, PI90, SW94, CH95, PI97}. The existence of the Li dip \citep{WA65, BT86} for stars between 1.1 $M_{\sun}$ and 1.5 $M_{\sun}$, with the exact range being dependent upon metallicity \citep{AT21}, 
as well as the increasing degree of depletion of Li for stars with decreasing mass at a rate significantly higher than predicted, clearly implies a missing physical process to drive the mixing in these stars. The growing observational evidence indicates that while rotation matters, it is the rate of spindown among stars that drives mixing, both for stars above the Li dip \citep{DE19, TW20, AT21}, and early on in the evolution of lower mass stars \citep{AT18, JE21, SU22}. Because the Li dip is well formed by the age of the Hyades, the implication is that the stars occupying this mass range have spun down to reduced rotation speeds compared to their ZAMS values by 650 million years, occupying what is known as the Kraft curve \citep{KR67} in rotation velocity. Beyond the age of the Hyades, these stars continue to spin down, though at a decreased rate, as illustrated by the comparison of NGC 752 with the Hyades \citep{BO22} and analysis of the even older cluster, NGC 6819 \citep{DE19}. 

Outside the Li dip toward lower masses, i.e. below 1.2 $M_{\sun}$, the Li plateau is defined by stars which initially, i.e. by the age of the Hyades, have little Li depletion (0.1 dex) compared to the primordial Hyades value. As these stars age, the plateau remains, but the level gradually declines. By the age of NGC 752, the plateau is approximately 0.3 dex below that defined by the Hyades at the same mass \citep{BO22}. The decline in the plateau level between the Hyades and NGC 752 is beautifully matched by the decline in $v$ sin$i$ between the Hyades stars and those in NGC 752 \citep{BO22}. What is surprising is that beyond the age of NGC 752, within the plateau in similar to much older clusters like IC 4651 (1.5 Gyr) \citep{AT00, AT09}, NGC 3680 (1.8 Gyr) \citep{AT09}, NGC 6819 (2.25 Gyr) \citep{AT14, DE19} and M67 \citep{CU17}, the Li level is the same as that for NGC 752 (IC 4651, NGC 3680, NGC 6819) or slightly (0.15 dex) lower (M67), within the uncertainties. Even the super-metal-rich cluster, NGC 6253 \citep{AT09}, where the stars populating the turnoff and subgiant branch come from a higher-mass Li dip due to supermetallicity, has a plateau level identical with that of M67 \citep{CU17} at an age of $\sim$3.0 Gyr. For completeness, subtle secondary effects may play a role in producing the apparent convergence of the Li plateau with increasing age. For example, increased metallicity shifts the the mass range for the Li-dip to higher masses, placing stars of higher mass and potentially higher initial rotation rates at the plateau. This effect may be counterbalanced by the observation that Galactic Li production leads to a higher initial value of A(Li) for young clusters with higher [Fe/H] \citep{CU11}.  

If mixing and Li depletion are driven by the rotational decline of stars on the plateau, the simple interpretation is that the degree of spindown for stars just cooler than the Li dip is so small after the age of NGC 752 that rotationally induced mixing becomes small, though just how small remains an open question. Since the primary star in 11028 falls in exactly the mass range where the plateau appears, it is probable that the stars populating the blue limit of the turnoff below the HEP have atmospheres that have remained stable, i.e. subject to little or no mixing, for the last 1-2 Gyr. The stars more massive than the plateau, those occupying the HEP and post-HEP, come directly from the Li dip, but the Li dip was in place by the age of the Hyades. If these stars reached their minimum rotational velocity by this age, it is likely that these stars, despite obvious evidence for significant mixing at an earlier main sequence age, remain stable to further mixing even longer than the stars in the plateau and until they reach the subgiant branch.

The final focus of the discussion are the seven filled blue circles in Figure 8. These stars (1020, 3050, 7026, 7044, 8006, 8048, 10055) have been tagged as unusual because of their locations in the CMD blueward of the ZAMS or mean turnoff relation. While the separation of the four brighter stars is unarguable, it should be emphasized that the fainter three points exhibit the same blueward positions independently in both the $V, (b-y)$ and $G, (B_{R}-B_{P})$ diagrams. All the stars have been included in the long-term radial-velocity survey \citep{GE15} and all have radial-velocity membership probabilities between 95\% and 98\%. None of the stars falls within the photometric binary category. In fact, most lie either on the ZAMS or below it, within the uncertainties. 

The common assumption for the existence of stars that lie blueward and brighter than the turnoff of any cluster is that they are blue stragglers (BS), especially for a cluster like M67 which has long been known for its rich BS population. The commonly accepted scenario for the formation of such systems is a binary mass transfer/merger event that turns the lower mass companion into a higher mass star, potentially leaving a visually faint but UV-bright white dwarf companion behind \citep{MC64, HI76, LE89, PE09}. The literature in support of this scenario is heavily tied to M67 and growing (see, e.g. \citet{LE19, SU20, PA21, LE21, GE21} and the many references therein.) The high percentage of binaries among BS in M67 is detailed in \citet{GE15}, supplying an obvious source for most photometric and/or spectroscopic anomalies. Two alternative means of identifying BS systems, particularly where the radial-velocity variations may be small to negligible,  make use of the presence of a UV bright white dwarf companion \citep{SI18} or, more recently, searching for stars with anomalously high rotation rates, indicative of a spinup caused by mass-transfer/merger \citep{LE19}. These two approaches should be relevant for systems that fall outside the CMD zones canonically occupied by BS, e.g. bluer than but fainter than the turnoff and recently formed BS still on the main sequence. The latter class of stars has been named {\it blue lurkers} \citep{LE19}.

As already noted, all radial-velocity binary and/or BS systems identified in \citet{GE15} have been excluded from our analysis so detection of a BS binary origin for the seven bluer stars of Figure 8 must come from an alternative technique; the majority of M67 BS systems have been detected in the UV via their white dwarf companion \citep{PA21}. We have searched the primary sources for matches to our seven anomalous stars, the UVIT Catalog of Open Clusters \citep{JA21} and the potential blue lurkers identified in M67 \citep{LE19, JA19, SU20}. Unfortunately, of the seven stars, only two lie within the UVIT field for M67, 1020 and 8006. The brightest of the blue stars in Figure 8, 1020, is classified as a blue lurker \citep{LE19}. While not identified as a blue lurker, 8006, the bluest star in Figure 8, is one of the 25 brightest stars detected in the far-UV in the field of M67 and is actually brighter than 1020 at these wavelengths. From SED analysis, \citet{JA19} find it probable that 8006 has a low mass white dwarf companion but 
indicate that further observations are required to make this claim definitive. It seems highly probable that these two stars are traditional binary BS with white dwarf companions.

To get some possible insight into the remaining five stars, we can examine the stars classed as single-star blue lurkers. In addition to 1020, \citet{LE19} also list 2001 and 7035, stars that have rotational periods of 5.6 and 8.0 days, respectively. 2001 isn't included in our final sample because it was excluded for astrometric reasons; even taking its astrometric uncertainty into account, the star sits 5$\sigma$ away from the cluster mean $\mu_{\alpha}$. For 7035, its membership is not in question. It was excluded from the discussion of the turnoff region because it had been classified as a photometric variable \citep{GE15}. If we insert both stars into the sample, 7035 is essentially a photometric twin for our anomalous blue candidate, 7044. In Figure 8, 7035 sits at ($(b-y)_0$, $M_{V}$) = (0.324, 3.575), leading to the possibility that 7044 is a likely blue lurker. The situation with 2001 is more complex. It sits at the red edge of the HEP, with ($(b-y)_0$, $M_{V}$) = (0.358, 3.471). While 11008 
occupies almost the identical position in the CMD, it isn't a photometric twin in one key respect. The $c_1$ index for 2001 is lower by 0.046 mag compared to 11008. This is crucial because ($\delta$$c_{1}$, $\delta$$V$) becomes (0.0, 0.8), placing the star solidly within the binary category of Figure 5. If it is a binary, any white dwarf 
companion cannot be the source of the extra luminosity.  Whether the CMD position is an indication of 
nonmembership, true binarity, or a side effect of its rapid rotation (rapid rotators can support the same mass at a lower/redder temperature/color) remains an open question. 

The final single-star member classed as a possible blue lurker is 11005 \citep{SU20}. This star was detected within the UVIT Catalog, but only in the far-UV. SED analysis 
indicates the presence of a low-mass white dwarf companion but, surprisingly, the star is a slow rotator. Its position in the CMD of Figure 8 is ($(b-y)_0$, $M_{V}$) = (0.335, 2.991), i.e. on the subgiant branch, which might explain the slow rotation. It's $c_{1}$ index implies a single star.

\section{Conclusions}
The open cluster, M67, is a rich and rapidly growing source of insight on the evolution of stars of lower mass, with particular impact constraining models of the sun. The 
valuable insight which the cluster provides is intricately tied to the accuracy and precision of the parameters that define the cluster, particularly the reddening, metallicity, distance, and age. While there has been a convergence of these values from an initial range of even 30 years ago, there is enough uncertainty, for example, to cast doubt on the suitability of using the cluster as a proxy for stars of solar composition and age, with some studies deriving supersolar metallicity and generally subsolar ages.

The $uvby$H$\beta$ photometry detailed in this investigation supplies a second wide-field, deep set of precision data capable of being used as photometric standards and 
tied directly to the extensive CCD data for NGC 752 (Paper I), covering stars over a wide range of temperature and luminosity. The calibration and comparison of the two clusters demonstrates that M67 is definitely more metal-rich than NGC 752, as defined by either $m_1$ or $hk$ data, by approximately 0.062 dex. Depending upon the choice of the Hyades metallicity (+0.12 or +0.15) \citep{CU17}, M67 has either [Fe/H] = +0.03 or +0.06. It should be emphasized that these values are defined by the F and early G turnoff stars. Since both indices lose sensitivity to metallicity changes near solar metallicity among giants, it is impossible to test if the giants are more metal-rich than the turnoff stars, as claimed if significant diffusion operates on the turnoff stars. It should be noted that the lack of metallicity sensitivity does allow the red giants to be used as a test of the differential reddening estimate between the clusters, since both clump stars and first-ascent red giants at a given ($(b-y)$) should have the same $hk$. While the sample of NGC 752 red giants is small in size and color range, comparison with M67 nicely constrains the derived offset ($\delta$$E(b-y) = -0.005$) from the few hundred stars on the main sequence, i.e. $E(b-y)$ is smaller in M67 than NGC 752 by no more than 0.010 mag. Returning to the turnoff metallicity, it is impossible to say at present what the impact is of diffusion on the specific photometric intermediate and narrow-band indices, though it is likely that the $hk$ index is directly tied to [Ca/H] while $m_1$ feels a broader impact from the overall change in [m/H].  

With the reddening and metallicity in hand, the cluster age and distance should be readily determinable. Before doing this, however, {\it Gaia} DR3 astrometry, in conjunction with the exquisite 40-year radial-velocity database \citep{GE15}, is used to eliminate probable nonmembers and binaries. This tightly constrained sample of single-star members with the highest precision parallaxes leads to an absolute distance modulus of $(m-M)_0$ = 9.69 and, combined with the reddening, $(m-M)$ = 9.77 $\pm$ 0.03 (sem), where the uncertainty is dominated by the small uncertainty in the reddening, Again, it should be emphasized that the change in the mean parallax from earlier work is consistent with the claims of a zero-point error in the early parallaxes but assumes that none exists in the current database.

With radial-velocity binaries removed, the precision photometry of the stars at the turnoff can be used to identify potential binary systems composed of stars of comparable mass that may have been missed by radial-velocity analysis, especially in the CMD region where the binary sequence normally crosses the vertical turnoff. With all known potential sources of scatter eliminated, the parallax-based distances for the clusters can be tested against isochrones of the appropriate metallicity, leading to a color-dependent correction for the coolest stars on the ZAMS, but an excellent match otherwise for $(m-M)$ = 8.33 and 1.45 Gyr for the apparent distance and age of NGC 752, respectively. For M67, using the same color-dependent correction defined by NGC 752, an excellent match to the ZAMS and subgiant branch for an age of 3.7 Gyr and apparent distance of $(m-M)$ = 9.77 is derived. The morphology of the turnoff, however, cannot be meshed with a single isochrone. Based upon the dispersion in color at a fixed luminosity, the age can be constrained with an uncertainty of $\pm$0.1 Gyr. However, the color at the bluest point below the HEP implies an age between 3.3 and 3.5 Gyr. Attempts to shift the data by tweaking the reddening, metallicity, and/or distance fail to supply an internally consistent solution for the three key parameters. As illustrated by the expansive discussion of the binary 11028 by SA and the broad-band analysis of M67 by Paper II, this problem reflects a fundamental discrepancy between the theoretical physics and the observations of stars near 1.2 $M_{\sun}$ in M67. Whatever the source(s) of the morphological distortions in the 
isochrones, they impact the entire turnoff region from the HEP to the Li plateau, i.e. to stars beyond the Li dip and potentially to the those of solar mass.

Finally, despite the removal of a wide array of photometrically and spectroscopically anomalous stars, a few still remain at the turnoff, blueward of the main sequence by amounts that appear unlikely to be caused by photometric scatter.  The blueward position is critical since combinations of stars on the normal CMD track cannot create a composite bluer than either star. So, unless the system is a composite of a normal turnoff star and a UV-bright source like a white dwarf, as appears to be the case for three of the anomalous stars, the remaining systems cannot be simple binaries. If they aren't mass transfer/merger systems, their evolution off the ZAMS must have been delayed by some physical process that occurs in a more extreme form than for the vast majority of stars evolving off the main sequence and toward the HEP. It is possible that this phenomenon is related to the distortion that more extensively impacts the broad CMD distribution of stars at the turnoff relative to the isochrones.
 
\acknowledgments
NSF support for this project was provided to BJAT and BAT through NSF grant AST-1211621, and to CPD through NSF grants AST-1211699 and AST-1909456. The filters used in the program were obtained by BJAT and BAT through NSF grant AST-0321247 to the University of Kansas. Extensive use was made of the WEBDA database maintained by E. Paunzen at the University of Vienna, Austria (http://www.univie.ac.at/webda). It is a pleasure to thank Eric Sandquist for making a draft copy of the NGC 752 eclipsing binary analysis available during revisions of this manuscript. It is a pleasure to thank the referee whose careful reading of the manuscript led to changes that enhanced the clarity of the discussion.

This work has made use of data from the European Space Agency (ESA) mission {\it Gaia}, processed by the Gaia Data Processing and Analysis Consortium (DPAC). Funding for the DPAC has been provided by national institutions, in particular the institutions participating in the Gaia Multilateral Agreement.

\facility{WIYN: 0.9m}

\software{IRAF \citet{TODY}}

\end{document}